\documentclass[preprint,final]{aastex}
\shorttitle{BVI Variability Survey of M3}
\shortauthors{Hartman et al.}
\begin{document}

\title{BVI Photometric Variability Survey of M3.}

\author{J.~D.~Hartman\altaffilmark{1}, J.~Kaluzny\altaffilmark{2}, A.~Szentgyorgyi\altaffilmark{1} \& K.~Z.~Stanek\altaffilmark{1}}
\altaffiltext{1}{Harvard-Smithsonian Center for Astrophysics, 60 Garden St., Cambridge, MA~02138, USA}
\altaffiltext{2}{Nicolaus Copernicus Astronomical Center, ul.~Bartycka 18, 00-716 Warszawa, Poland}
\email{jhartman@cfa.harvard.edu, jka@camk.edu.pl, aszentgyorgyi, kstanek@cfa.harvard.edu}

\begin{abstract}
We have conducted a three band ($BVI$) variability survey of the
globular cluster M3. This is the first three band survey of the
cluster using modern image subtraction techniques. Observations were
made over 9 nights in 1998 on the 1.2~m.~telescope at the
F.~L.~Whipple Observatory in Arizona. We present photometry for 180
variable stars in the M3 field, of which 12 are newly discovered. New
discoveries include six SX Phe type variables which all lie in the
blue straggler region of the color magnitude diagram, two new first
overtone RR Lyrae, a candidate multi-mode RR Lyrae, a detached eclipsing binary, and two unclassified
variables. We also provide revised periods for 52 of the 168
previously known variables that we observe. The catalog and photometry
for the variable stars are available via anonymous ftp at
ftp://cfa-ftp.harvard.edu/pub/kstanek/M3/.
\end{abstract}

\section{Introduction}

The globular cluster M3 (NGC 5272) is one of the most studied clusters in the
Galaxy. Since the work of Bailey\ (1913) and Shapley\ (1914), there have been numerous
surveys for photometric variability in the cluster. In 1973, Sawyer-Hogg compiled the most complete catalog of variables in the cluster prior to the modern era. Recently Bakos,
Benko, \& Jurcsik\ (2000) compiled a catalog of 274 variable stars in
the cluster, and published improved identification and astrometry for
them. M3 is
particularly known for containing the largest number of RR Lyrae
variables, with more than 182 such stars (Clement et al.\ 2001). Many
groups have focused on studying the population of RR Lyrae
variables. Corwin et al.\ (2001) and Clementini et al.\ (2004) have
published a catalog of 222 confirmed or suspected RR Lyrae variables,
and have provided a detailed analysis of the 8 known double-mode RR
Lyraes (RRd). Other surveys have focused specifically on searches for variability in blue straggler stars (BSS), notably Kaluzny et al.\ (1998) discovered only one SX Phe star out of 25 monitored BSS. They also discovered a contact binary star near the base of the red giant branch.

In this contribution we present the data from a $BVI$ variability survey
of the cluster performed in the spring of 1998. This is the first
three band survey of the cluster using modern image subtraction
techniques. New results include the discovery of 6 new SX Phe
variables, 2 new first-overtone RR Lyraes, a candidate multi-mode RR Lyrae, a detached eclipsing binary, and 2 new variables that we do not classify. We also provide
revised periods for 52 of the 168 previously known variables that we
observe.

We describe our observations and data reduction in the following
section. In \S 3 we present the catalog of variables, in \S 4 we analyze the light curve of the eclipsing binary NV296, in \S 5 we briefly
describe the population of variables using color magnitude diagrams,
and in \S 6 we discuss the results.

\section{Observations and Data Reduction}

\begin{figure}[p]
\epsscale{1}
\plotone{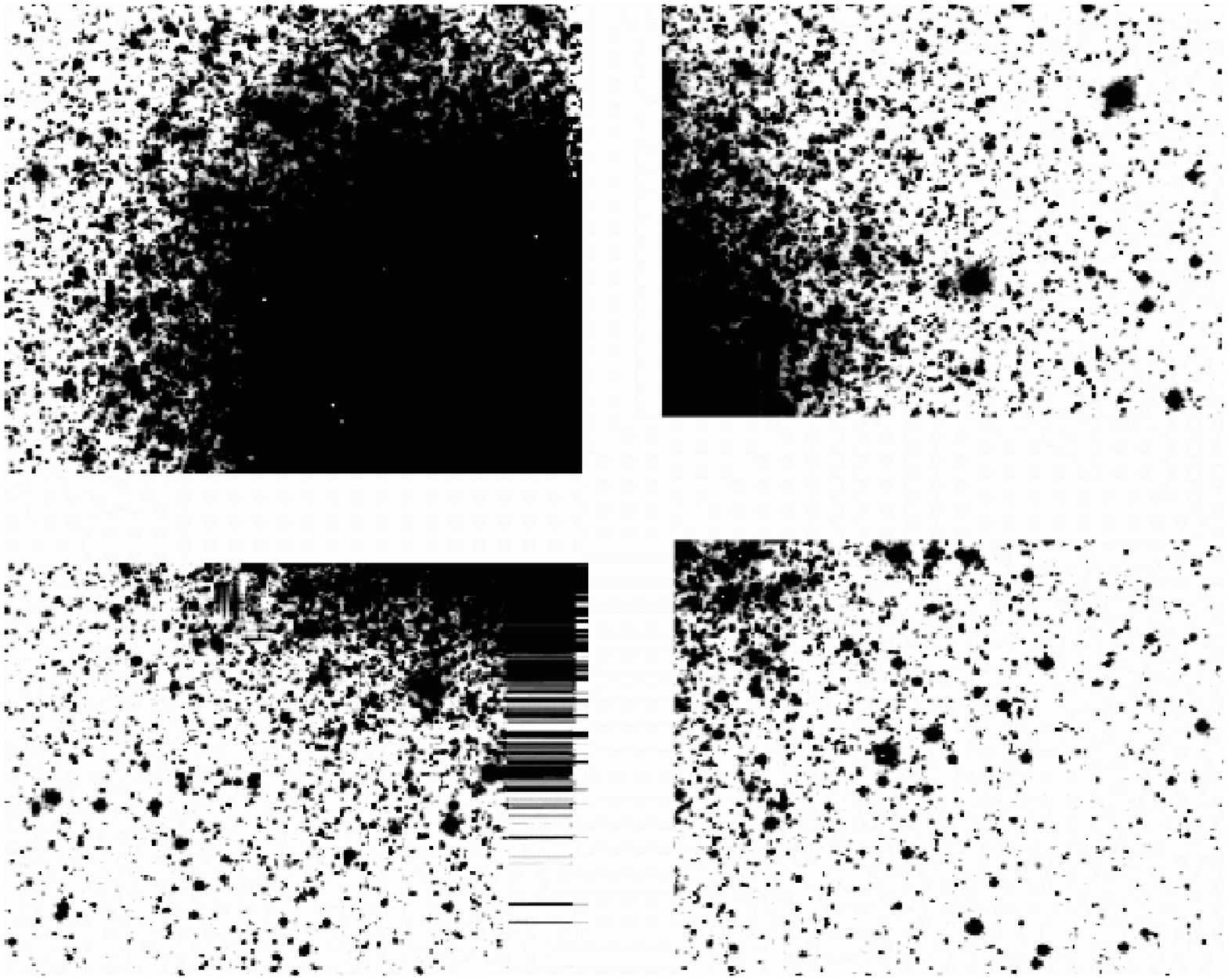}
\caption{A mosaic of the four chips showing the position of M3 within the 18$\arcmin$x18$\arcmin$ field of view. The horizontal lines in chip 2 are the result of running fixpix across a number of bad columns.}
\label{fov}
\end{figure}

The observations were made at the F.~L.~Whipple Observatory (FLWO)
1.2~m telescope using the 4-Shooter CCD mosaic containing four
thinned, back-side-illuminated, AR-coated Loral 2048x2048 CCDs. The
camera has a pixel scale of $0\farcs333$~pixel$^{-1}$ and a field of
view (FOV) of $11\farcm4\times 11\farcm4$ for each chip. We obtained
observations on 9 nights between the dates of April 16-20, 1998 and May
28-31, 1998. These observations consist of 4~x~300~s, and 67~x~420~s
$B$-band exposures, 14~x~240~s, 198~x~300~s, and 13~x~450~s $V$-band
exposures and 4~x~180~s, and 72~x~240~s $I$-band exposures. The
sampling cadence was as low as 7 minutes, allowing for the detection
of very short period SX Phe variables. The center of M3 at
$(\alpha,\delta)=(13^{\rm h}42^{\rm
m}11\fs2,+28\arcdeg22\arcmin32\arcsec)$ (J2000.0) was positioned in
the lower, right corner of chip 3. A mosaic of all four chips is shown in
Fig~\ref{fov}.

The preliminary CCD reductions were performed using the standard
routines in the IRAF CCDPROC package\footnote{IRAF is distributed by
the National Optical Astronomy Observatories, which is operated by the
Association of Universities for Research in Astronomy, Inc., under
agreement with the National Science Foundation.}.

To obtain photometry we used the image subtraction techniques due to
Alard \& Lupton\ (1998; also Alard\ 2000) as implemented in the ISIS
2.1\footnote{ISIS package is available from C. Alard's web-site at
http://www2.iap.fr/users/alard/package.html} package. To detect variables we first took the absolute value of all the subtracted images and stacked them to form a single image. We then searched this image for strong point sources, identifying these as variables. We obtained
light curves only for sources identified as variable in this way by performing weighted-aperture photometry on the subtracted images. This
procedure provides differential photometry in Analog-to-Digital Units (ADUs), to convert to
magnitudes we obtained profile photometry for the stars in our field
using the DAOPHOT/ALLSTAR package (Stetson\ 1987, 1991). As a sanity
check, we also generated light curves using profile photometry which
resulted in generally worse precision and yielded no additional
variables. The method that we have just described is the same
technique used by Kaluzny et al.\ (2001); the procedure is discussed
in more detail in that paper.

We also generated light curves for blue straggler stars, examining them individually for variability. We identified a handful of low-amplitude ($0.04$ mag) SX Phe variables in this fashion.

To transform our light curves to standard BVI magnitudes we observed a
total of 128 stars from 22 Landolt\ (1992) fields so that the
transformation for each chip utilized typically 30 stars from 5
fields. The airmass of these observations ranged from 1.08 to
1.93. The resulting uncertainty in the zero-point of our observations
is 0.02 magnitudes. As a consistency check we calculated the median
colors for the turnoff stars independently for each chip. The results
shown in Table~\ref{tab:tab1} are consistent with an uncertainty of
0.02 magnitudes.

\begin{figure}[p]
\epsscale{1}
\plotone{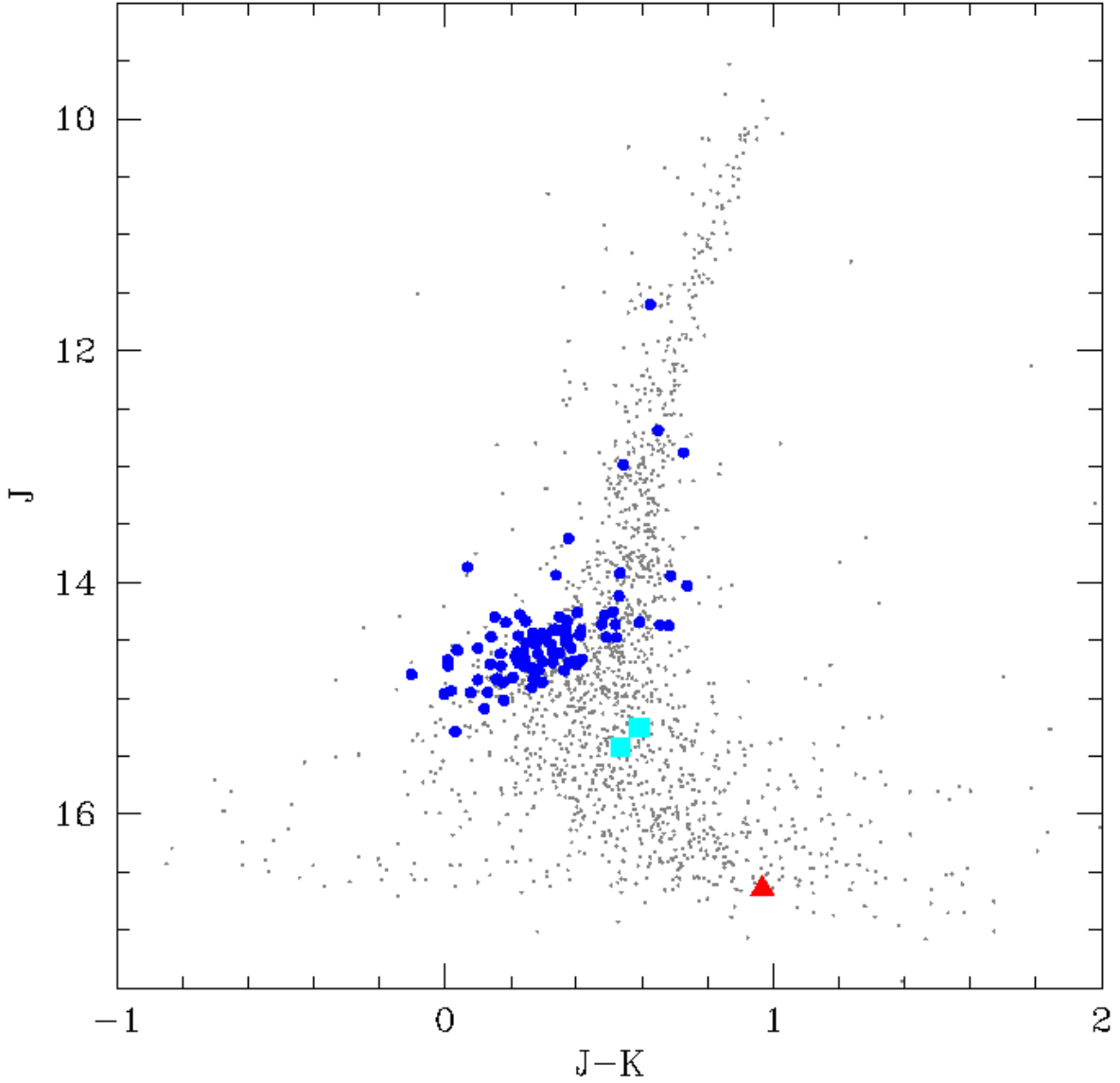}
\caption{{\it 2MASS}\/ $J$ vs. $J-K$ CMD for the globular cluster M3. The filled circles denote {\it 2MASS}\/ sources that match to RR Lyrae in our catalog, filled squares denote sources that match to unclassified variables in our catalog, and the filled triangle shows the source that matched to an SX Phe variable in our catalog. The fact that the RR Lyrae cluster in a single region of the CMD confirms that these matches to {\it 2MASS}\/ are real.}
\label{2mass}
\end{figure}

\begin{figure}[p]
\epsscale{1}
\plotone{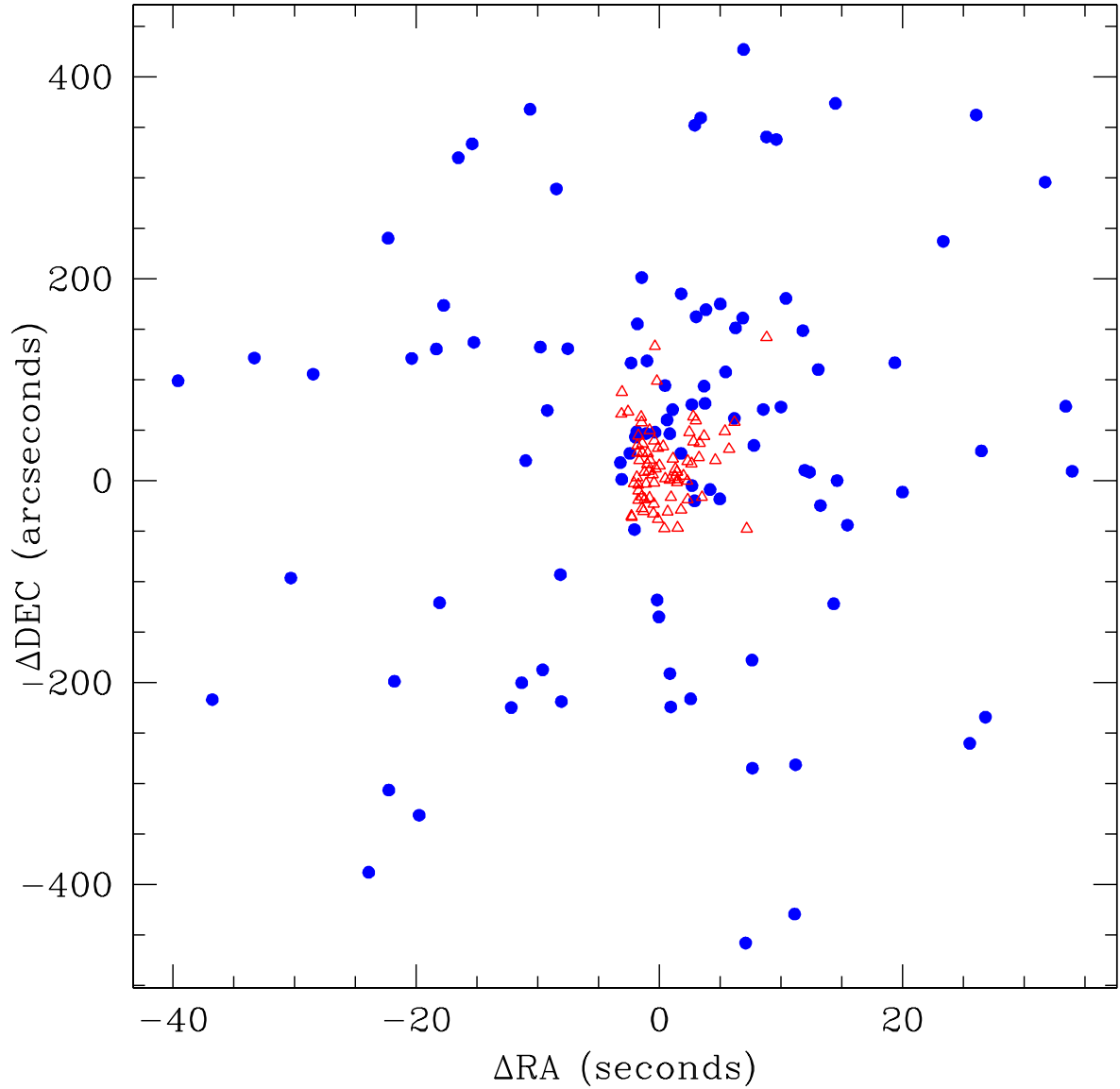}
\caption{Distance from the center of the cluster of RR Lyrae that had matches to {\it 2MASS}\/ (filled circles) as well as those that did not (open triangles). From this it is clear that non-matches are the result of incompleteness in the {\it 2MASS}\/ catalog in the crowded center of the cluster.}
\label{RRLyr_location}
\end{figure}

We obtained astrometry for the variables by first matching to the catalog of
Bakos et al.\ (2000). We used at least 10 preliminary matches on each
chip to obtain the transformation between rectangular and equatorial
coordinates. The residuals of the matched stars were less than
$0\farcs1$. Using the RA/DEC from this transformation we proceded to match our sources to the Two Micron All Sky Survey ({\it 2MASS}\/; Skrutskie et al.\ 1997) point-source catalog using a matching radius of $0\farcs7$ for a total of 98 matches. The median residual was $0\farcs43$. All but three of these matches are RR Lyrae. Both of the unclassified newly discovered variables and one of the newly discovered SX Phe variables matched to sources in {\it 2MASS}\/. Figure~\ref{2mass} shows the location of 93 of the matched sources on the {\it 2MASS}\/ $J$ vs. $J-K$ color-magnitude-diagram (CMD) for M3. The five matched sources not shown in this figure were missing $K$ photometry. The fact that all but a few sources lie clumped near a region of constant $J$ (the horizontal branch) suggests that these matches to RR Lyrae are correct, with the scatter likely due to variability of the RR Lyrae and blending with nearby sources. In Figure~\ref{RRLyr_location} we plot the position on the sky of all RR Lyrae in our catalog, together with the positions of those that match to {\it 2MASS}\/. From this figure it is clear that the non-matches are due to incompleteness in the {\it 2MASS}\/ catalog near the crowded center of the globular cluster.

\section{Catalog of Variables}

Following the above procedures we identify 180 variable sources of
which 12 are newly identified as variables. The newly identified
variables include 6 new SX Phe stars, 2
first overtone RR Lyrae (RR1), a candidate multi-mode RR Lyrae (RR01), one detached eclipsing
binary (EB), and 2 unclassified variables. We also provide revised periods for 52 of the known
variables. The revisions were determined based on visual comparisons between the light curves phased with the published period and the light curves phased with the period calculated using the Schwarzenberg-Czerny\ (1996) algorithm. We consider the period to be necessary of revision if the published period is incompatible with the observed light curve given the formal errors from the photometry. We have recovered one of the previously identified SX Phe stars (V237), we note however that the coordinates given for this variable appear to have been switched with the coordinates for V238 in the discovery paper by Kaluzny et al.\ (1998), this error has been propagated through the catalogs of Bakos et al.\ (2000) and Clement et al.\ (2001), we also revise the period for this variable. Finding charts for the newly discovered variables are shown
in Figure~\ref{finder_chart}. A machine-readable version of the catalog, as well as $V$, $B$, $I$, $B-V$ and $V-I$ light curves for the variable stars are available via anonymous ftp\footnote{ftp://cfa-ftp.harvard.edu/pub/kstanek/M3/}. We present the catalog in
Tables~\ref{tab:tab2} and \ref{tab:tab3}, the columns are as follows:

\renewcommand{\labelitemi}{--}
\renewcommand{\labelitemii}{--}

\begin{itemize}
\item {\it ID}\/: The ID number of the variable. V1-V285 is taken from Clement et al.\ (2001). New identifications are denoted by ``NV.'' The ID number roughly corresponds to discovery order.
\item {\it 2MASS-flag}\/: Integer denoting whether or not the variable matched to a {\it 2MASS}\/ source. The flag is 0 for no match, and 1 for a match.
\item {\it RA}\/: Right ascension for epoch J2000.0. This value is taken from the {\it 2MASS}\/ catalog for sources with a {\it 2MASS}\/ match, otherwise it is obtained from the rectangular to equatorial transformation derived using the Bakos et al.\ (2000) catalog.
\item {\it DEC}\/: Declination for epoch J2000.0. This value is taken from the {\it 2MASS}\/ catalog for sources with a {\it 2MASS}\/ match, otherwise it is obtained from the rectangular to equatorial transformation derived using the Bakos et al.\ (2000) catalog.
\item {\it V$_{flag}$}\/: Integer denoting the type of V-band observations. Values are 1 for observations that could be converted to magnitudes or 2 for observations that were left in differential count units.
\item {\it B$_{flag}$}\/: Integer denoting the type of B-band observations. Values are 0 for no B-band detection, 1 for observations that could be converted to magnitudes, or 2 for observations that were left in differential count units.
\item {\it I$_{flag}$}\/: Integer denoting the type of I-band observations. Values are 0 for no I-band detection, 1 for observations that could be converted to magnitudes, or 2 for observations that were left in differential count units.
\item {\it A$_{V}$}\/: Observed full-amplitude in V-band, defined to be the faintest observed magnitude minus the brightest observed magnitude on the cleaned light curve.
\item $\langle V\rangle$\/: Flux averaged mean V magnitude of the star. For the eclipsing binary this is the out of eclipse magnitude determined with EBOP(see \S 4).
\item {\it A$_{B}$}\/: Observed full-amplitude in B-band, defined to be the faintest observed magnitude minus the brightest observed magnitude on the cleaned light curve.
\item $\langle B\rangle$\/: Flux averaged mean B magnitude of the star. For the eclipsing binary this is the out of eclipse magnitude determined with EBOP(see \S 4).
\item {\it A$_{I}$}\/: Observed full-amplitude in I-band, defined to be the faintest observed magnitude minus the brightest observed magnitude on the cleaned light curve.
\item $\langle I\rangle$\/: Flux averaged mean I magnitude of the star. For the eclipsing binary this is the out of eclipse magnitude determined with EBOP(see \S 4).
\item {\it Period}\/: The observed best period of variability in days. This was derived using the {\tt ANOVA} statistic of Schwarzenberg-Czerny\ (1996). In cases where aliasing allowed for a number of acceptable periods, we chose the period corresponding to the peak in the periodogram nearest to the published period, where available.
\item {\it Published Period}\/: The published period for the star taken from Clementini et al.\ (2004), Corwin \& Carney\ (2002) or Clement et al.\ (2001) in that order of priority.
\item {\it Period Revision Flag}\/: Integer denoting whether or not the observed best period should be taken as a revision of the published period. This determination is based on a visual comparison between the light curve phased with the published period and the light curve phased with the observed best period. Values are 1 for a revision, and 0 for no revision.
\item {\it JD of minimum V}\/: Julian Date (in 2450000) of the first minimum in V-band to occur after the first observation. This field is null for unclassified variables.
\item {\it Phase of minimum B}\/: Phase of the minimum in B-band, where 0 phase is set to the minimum in V-band.
\item {\it Phase of minimum I}\/: Phase of the minimum in I-band, where 0 phase is set to the minimum in V-band.
\item {\it Classification}\/: Classification of variability, the symbols are as follows:
\begin{itemize}
\item RR0: fundamental mode RR Lyr (RRab).
\item RR1: first overtone RR Lyr (RRc).
\item RR01: candidate multi-mode RR Lyr (RRd). Note that the possible presence of multiple periods was determined by eye and was not the result of a systematic search.
\item SXP: SX Phe type variable.
\item N/A: Unclassified variable.
\item EB: Eclipsing binary.
\end{itemize}
\item {\it Remarks}\/: Here we denote the possible presence of multiple modes (mp) for SXP variables as well as the presence of the Blazhko effect (Bl) within the observations.
\end{itemize}

Figure~\ref{lc} shows the $V$-band, $B-V$ and $V-I$ light curves for 12 of the 130 previously identified
variables in the catalog for which these 3 measurements are
available. We display $V$, $B-V$ and $V-I$ light curves, where available, for the 2 unclassified new variables and the 3 new RRs in Figure~\ref{LPVRR1} and light curves for the 6 new SX Phe variables in Figure~\ref{SXP}. The $BVI$ light curves for the eclipsing binary (NV297) are shown in Figure~\ref{EB} together with a model fit.

\section{Analysis of the Eclipsing Binary NV296}

Parameters for the eclipsing binary NV296 were determined using the Eclipsing Binary Orbit Program (EBOP)
model for detached eclipsing binaries (Nelson \& Davis\ 1972; Popper \& Etzel\ 1981) (see Figure~\ref{EB}). The first step was to obtain a fit to the V-band light curve. In doing so we varied the luminosity ratio, the radius of the primary, the inclination of the orbit, and the out-of-eclipse luminosity. We then obtained a second fit varying the ratio of the radii in place of the radius of the primary. We found empirically that the solution would not converge when the radius of the primary and the ratio of the radii were allowed to vary simultaneously. We held the mass ratio fixed at 1.0, and assumed that gravity darkening and the reflection effect were negligible. Because the period is relatively short ($0.445955$ days) we assumed that the orbits would have circularized, so we held the eccentricity fixed to 0. We note that when we allowed this parameter to vary the results were consistent with $e=0$. Having obtained a model for the V-band light curve, we proceeded to fit a model to the B-band light curve assuming the same orbital parameters and radii from the V-band model. We did not analyze the I-band light curve since noise and poor phase coverage during the eclipses conspired to prevent us from obtaining an acceptable fit.

Without spectra it is difficult to constrain
the spectral types and luminosity classes of the components, and hence the limb darkening
coefficients. As a preliminary analysis we fixed the limb darkening coefficients for both components to 0.57 in $V$ and 0.72 in $B$, which is consistent with an F2V star in the limb darkening tables from Claret \& Gimenez\ (1990). We caution that the results from this fit should be treated as preliminary until spectra can be obtained for this variable.. 

The parameters from the best fit model are shown in Table~\ref{Tab:EB}. To test the effect of the unconstrained mass ratio on the fit, we obtained an independent model assuming a mass ratio of 0.7. The parameters for this model are also shown in Table~\ref{Tab:EB}, we note that the results appear to be consistent, in particular the position of the components on the CMD is unaffected by the mass ratio. We find that the primary has $B-V=0.349 \pm 0.042$ mag, and $V=19.331 \pm 0.028$ mag, while the secondary has $B-V=0.395 \pm 0.064$ mag, and $V=19.747 \pm 0.042$ mag. The errors listed here and in Table~\ref{Tab:EB} are propagated from the standard errors produced by the EBOP fits, we caution that for all parameters except the color these errors are likely to be optimistic. In the case of the $B-V$ colors the errors listed are calculated assuming $B$ and $V$ are uncorrelated when in reality the two magnitudes will be correlated as a result of the fitting procedure. We note that for three different choices of the limb darkening coefficients $B-V$ changed by less than a percent for both components whereas the other parameters varied by amounts consistent with the standard errors.

\section{Color Magnitude Diagrams}

In Figures~\ref{cmbv} and ~\ref{cmvi} we plot the $B-V$ vs. $V$ and $V-I$
vs. $V$ CMDs for the cluster. In these figures
we use filled circles to show RR Lyr variables, open squares for SXP
variables, stars for unclassified variables and open triangles for
the two components of the EB. Blue straggler stars that were examined for variability are shown with filled triangles. Note that these stars were selected based on their position in the $B-V$ CMD. Where available, the locations of these stars in the $V-I$ CMD are also shown. 

The SXP variables for which colors could be obtained all lie
within the blue straggler region of the CMD as a result of our selection procedure. Although their location
on the CMD may indicate that these variables are indeed likely members
of the cluster, one must be cautious in such claims following the
identification with quasars of three variables in the M3 field lying
in the blue straggler region of the CMD (Meusinger, Sholz \& Irwin,
2001).

The components of the eclipsing binary NV296 appear to lie near the ZAMS at a distance of $(m-M)_V=15.04$ mag (Harris\ 1996). The photometry for the primary is consistent with an F2V dwarf, and the secondary is consistent with an F3-4V dwarf. If the eclipsing binary is indeed a member of the cluster then both components (particularly the primary) appear to lie slightly below the cluster main sequence, closer to the ZAMS for the cluster. 

The location of the unclassified variables on the CMD is completely uncertain since we have not observed a full period for either of these stars.

To further study the RR Lyr population we plot the horizontal branch
regions of the CMDs in detail in Figures~\ref{cmbv_hb} and
~\ref{cmvi_hb}. Here we distinguish between the types of RR Lyr
variables using filled circles for RRab, open circles for RRc and open
stars for RRd candidates. The evolution from RRab to
RRc along the horizontal branch is clearly demonstrated in the CMD for
M3. We note that some RR Lyrae may appear to lie outside of the instability strip as a result of uncertainties in their colors. We also note that blending near the center of the cluster may cause a few variables to appear artificially bright.

\section{Discussion}

There has been a great deal of interest in studying the variability of BSS in globular clusters. A number of globular clusters show a substantial population of variable blue stragglers. Gilliland et al.\ (1998) found that 9 out of 47 monitored BSS in 47 Tucanae showed variability above 0.02 mag. Recently Kaluzny et al\ (2004) identified 35 new SX Phe variables in $\omega$ Centauri. However, despite having a sizeable population of BSS, Kaluzny et al.\ (1998) found only a single SX Phe star in M3. We confirm the relative under-abundance of SX Phe stars in M3, finding only 7 variables out of 122 monitored BSS. We note that with image subtraction we would have been able to detect single-mode, short-period variability in the monitored BSS with amplitudes greater than 0.02 magnitudes (see for example NV289 in Figure~\ref{lc}). 

Eclipsing binaries are another interesting target for globular cluster variability surveys. Besides the possibility that binaries play an important role in the dynamical evolution of globular clusters (Hut et al.\ 1992) and the formation of the BSS (Leonard 1989; Leonard \& Fahlman 1991), eclipsing binaries can also be used as a tool to determine accurate ages and distances to the clusters (Paczy\'{n}ski 1997; see Kaluzny et al.\ 2004 for a discussion of a systematic program to search for detached eclipsing binaries in nearby globular clusters). The location of the components of NV296 on the CMD suggests that the binary may indeed be a member of the cluster. If it is a member, then there must be some mechanism that has allowed the components to avoid evolution into a giant. One hypothesis may be mass transfer onto the primary, however the fact that the light curve seems well fit by a detached eclipsing binary model suggests that mass transfer is not currently occurring at a significant rate. Moreover, the secondary star itself appears to be younger than other F3-4V stars in the cluster. It is necessary to obtain spectra for this system before anything more can be said about its membership. If it is a member, then spectra will be useful in determining the masses of the components.

As mentioned in the introduction, many of the recent studies of M3 have focused on the RR Lyrae population. Most recently Cacciari, Corwin \& Carney\ (2004) used data from previous $BVI$ surveys to conduct a detailed analysis of the RR Lyrae population in M3. Other authors have pointed to inconsistencies between existing models for horizontal branch evolution and the observed population (Catelan 2004; Clementini et al.\ 2004 discuss the observation of rapid evolution among the double mode RR Lyrae). Although we do not perform a systematic analysis of the RR Lyrae population, the data that we present should be useful for further investigations of these stars.

\acknowledgments{We gratefully acknowledge G.~Pojmanski for his excellent ``lc'' program, J.~Devor for his period-finding code, and G.~Torres for his help with the EBOP routine. This research has made use of the SIMBAD database, operated at CDS, Strasbourg, France. It has also made use of images from the Digitized Sky Survey which were produced at the Space Telescope Science Institute under U.S. Government grant NAG W-2166. The images of the DSS are based on photographic data obtained using the Oschin Schmidt Telescope on Palomar Mountain and the UK Schmidt Telescope. The plates were processed into the present compressed digital form with the permission of these institutions. JDH is funded by a National Science Foundation Graduate Student Research Fellowship.}

\newpage

\clearpage

\clearpage

\clearpage

\begin{figure}[p]
\epsscale{1}
\plotone{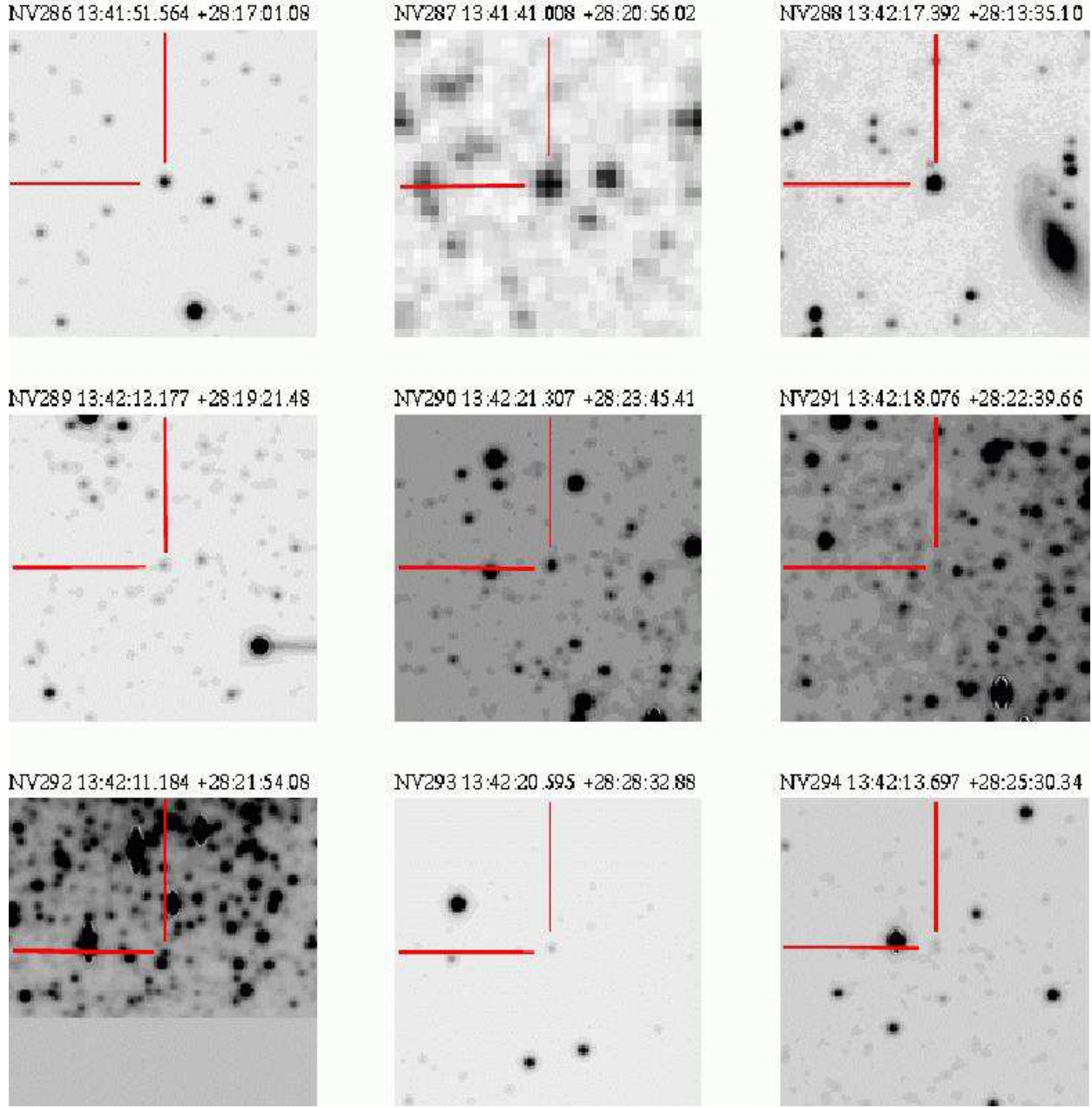}
\caption{1$\arcmin$x1$\arcmin$ finding charts for the 13 newly discovered variables. The chart for NV287 is from the DSS, all other images are taken from our observations.}
\label{finder_chart}
\end{figure}

\clearpage

\addtocounter{figure}{-1}
\begin{figure}[p]
\epsscale{1.0}
\plotone{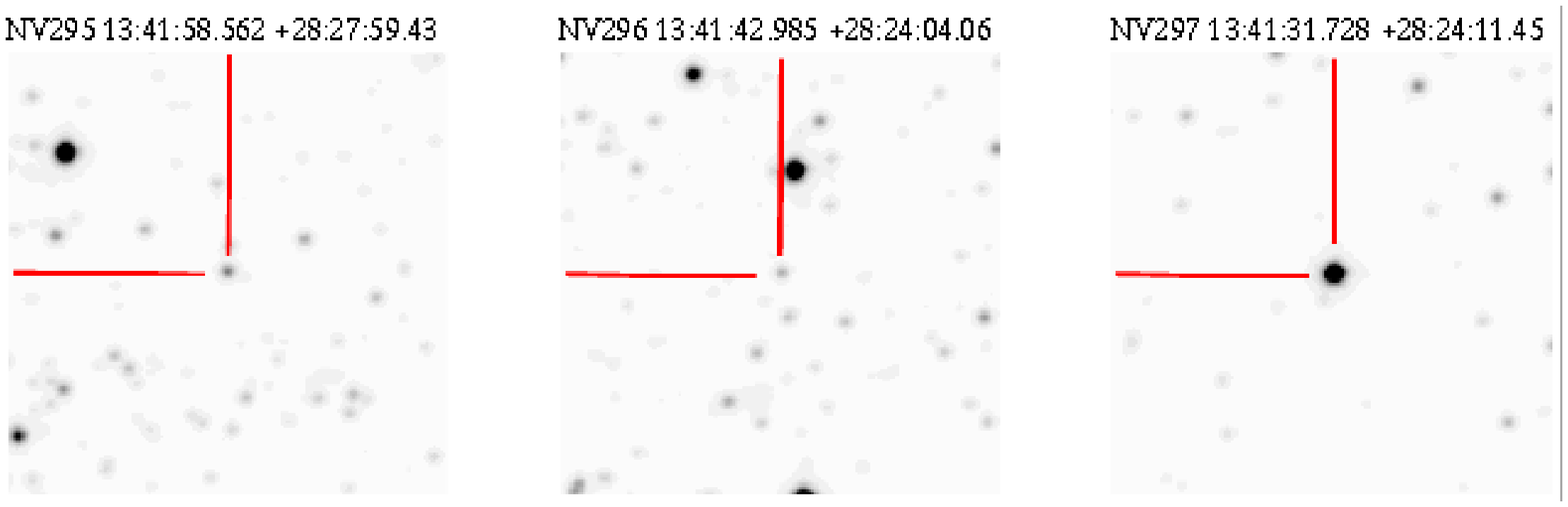}
\caption{{\it Continued}.}
\end{figure} 

\clearpage

\begin{figure}[p]
\epsscale{1}
\plotone{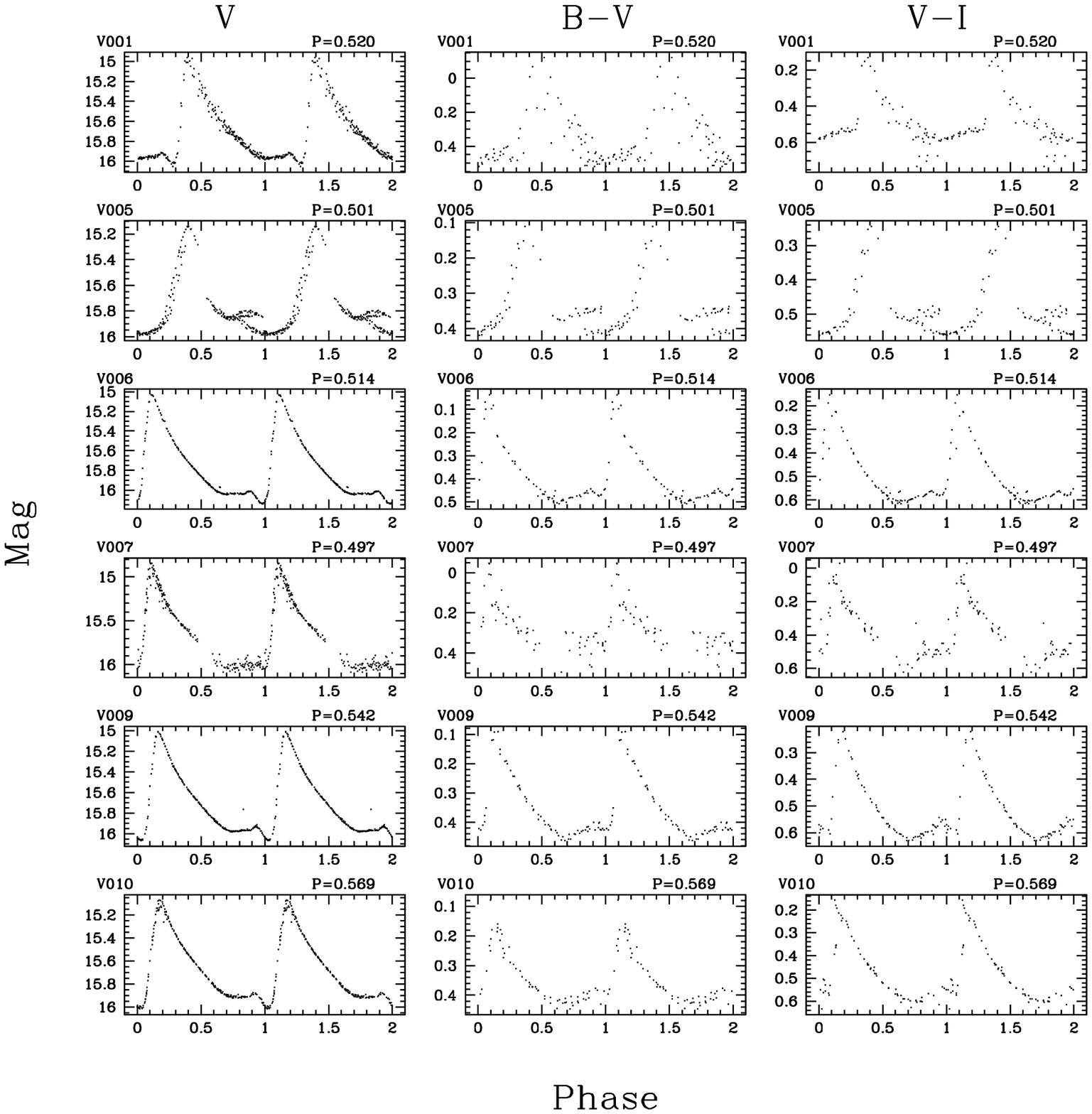}
\caption{V, B-V and V-I light curves for 12 of the 130 previously identified variables for which all three of these measurements were available. Phase = 0 is set to the minimum in V.}
\label{lc}
\end{figure}

\clearpage

\addtocounter{figure}{-1}
\begin{figure}[p]
\epsscale{1.0}
\plotone{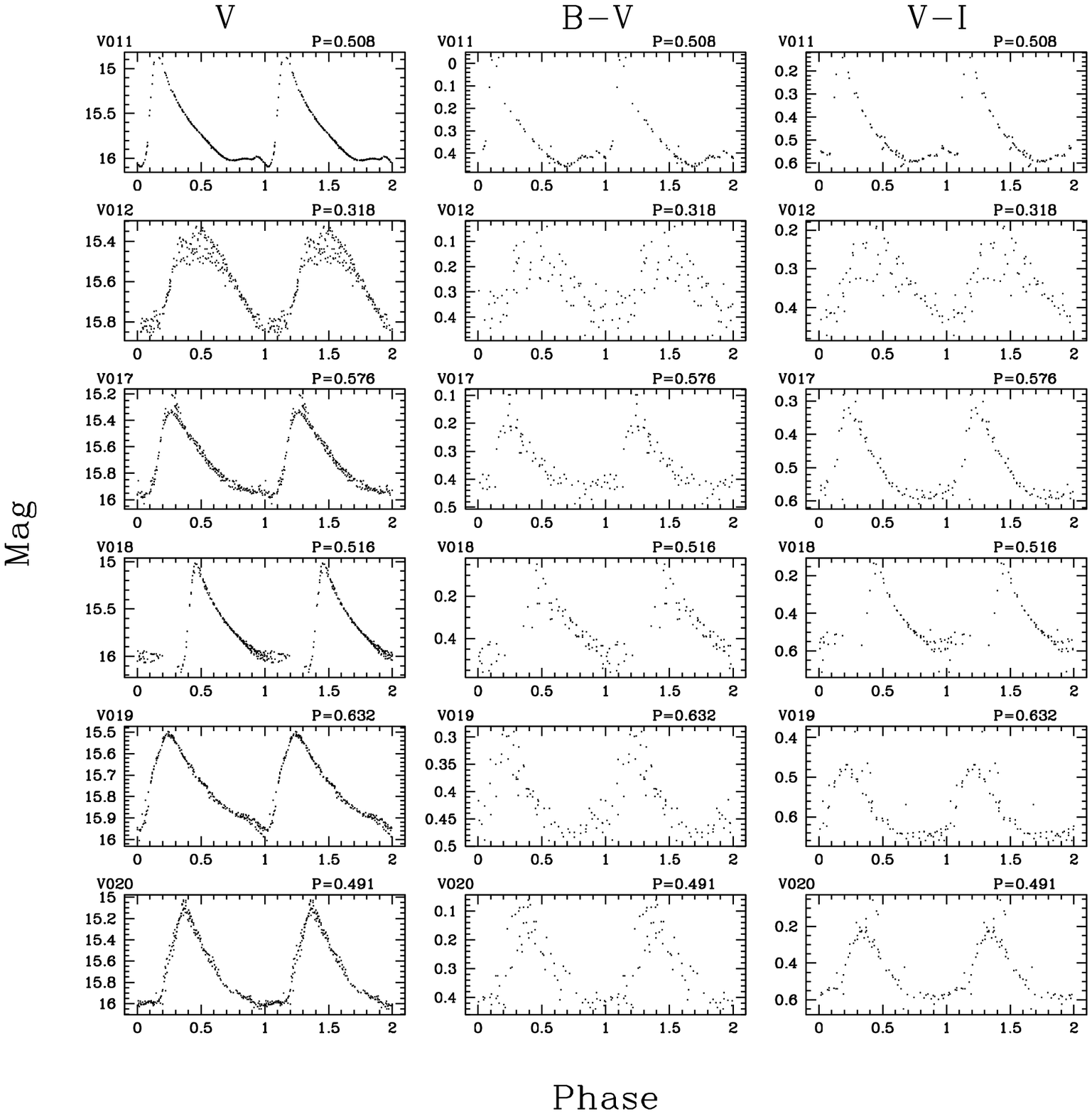}
\caption{{\it Continued}.}
\end{figure}

\clearpage

\begin{figure}[p]
\epsscale{1}
\plotone{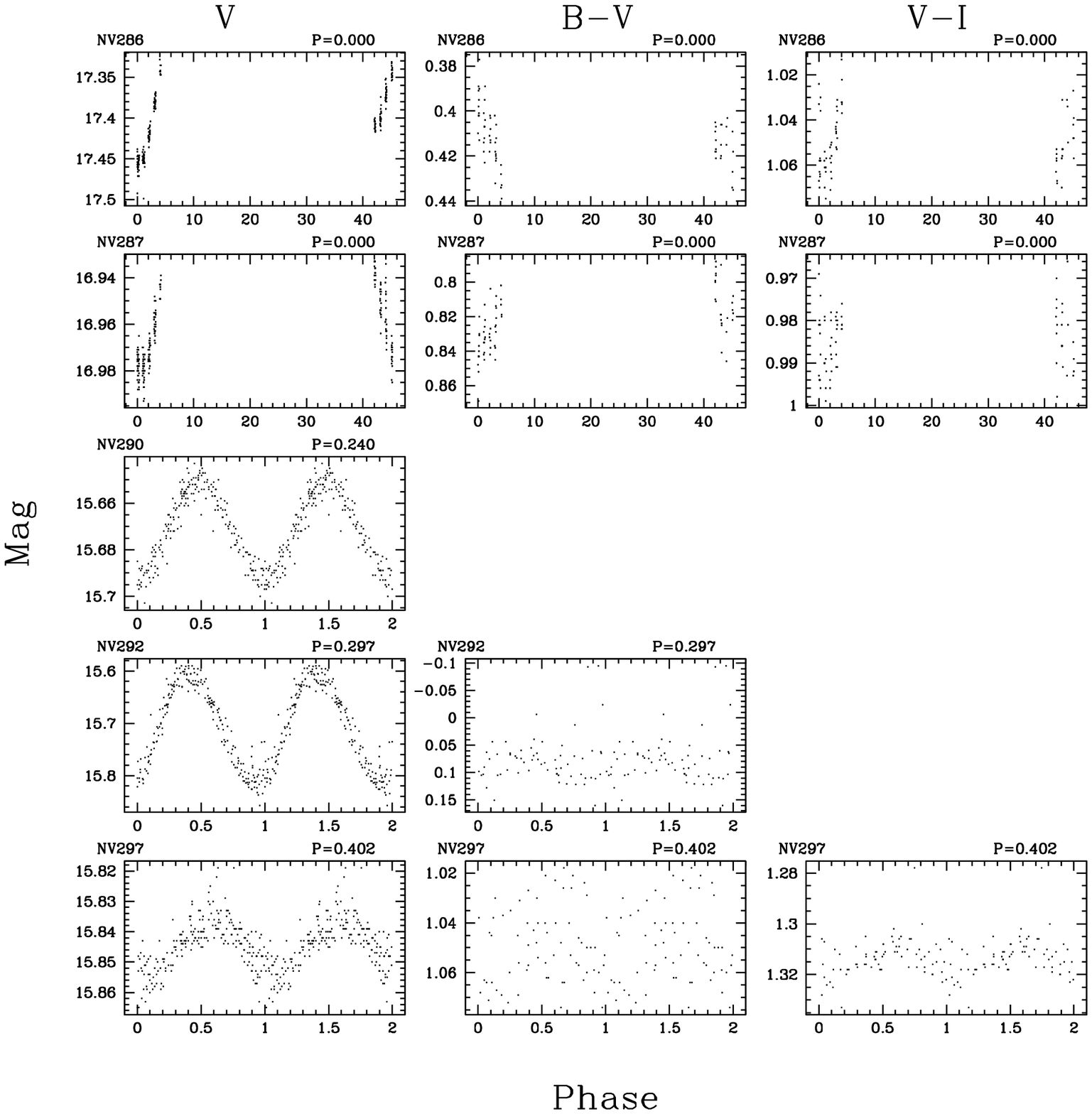}
\caption{$V$, $B-V$, and $V-I$ light curves for the two new unclassified variables and the three newly discovered RR variables.}
\label{LPVRR1}
\end{figure}

\clearpage

\begin{figure}[p]
\epsscale{1}
\plotone{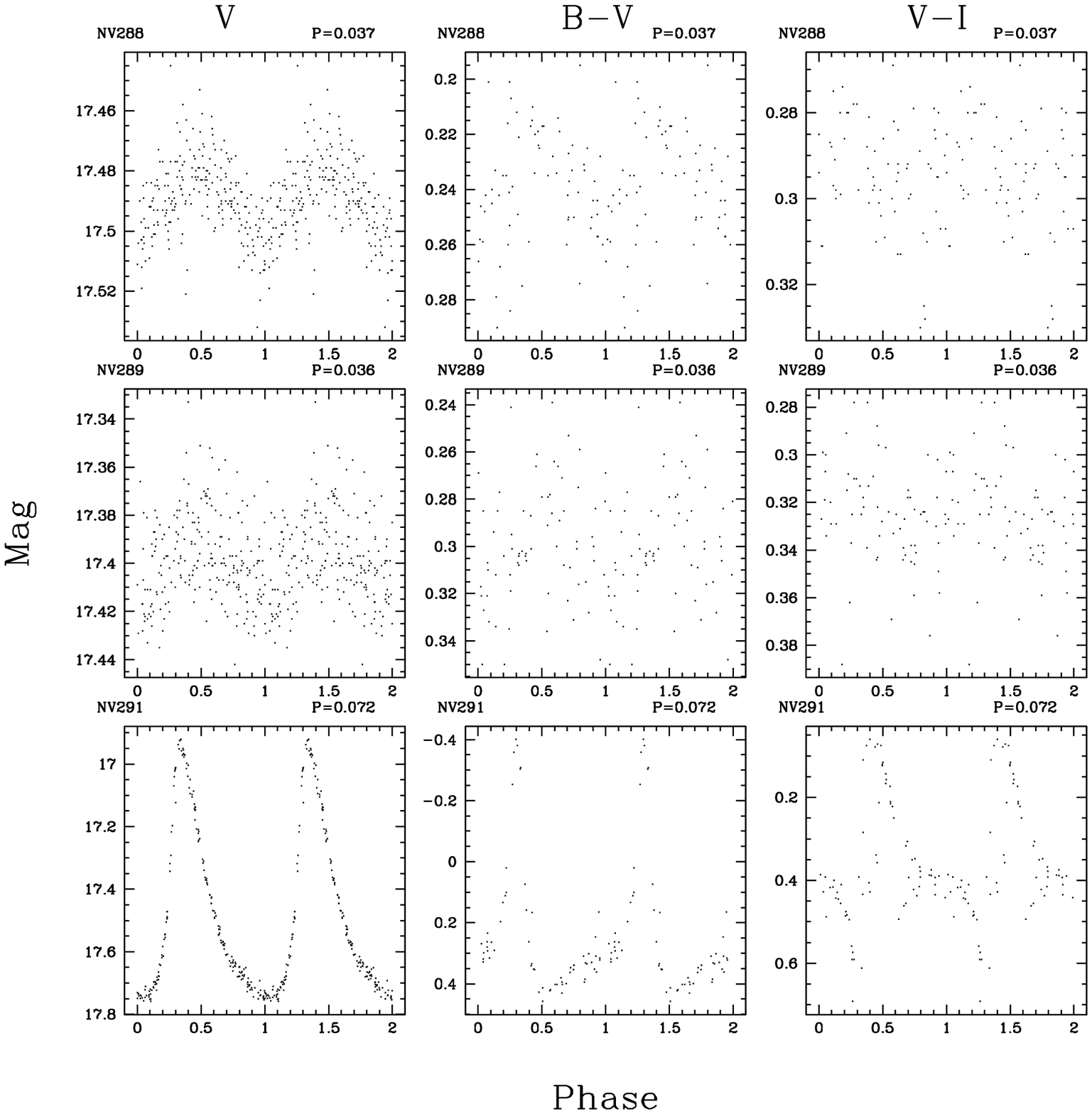}
\caption{$V$, $B-V$, and $V-I$ light curves for the six newly discovered SX Phe type variables.}
\label{SXP}
\end{figure}

\clearpage

\addtocounter{figure}{-1}
\begin{figure}[p]
\epsscale{1}
\plotone{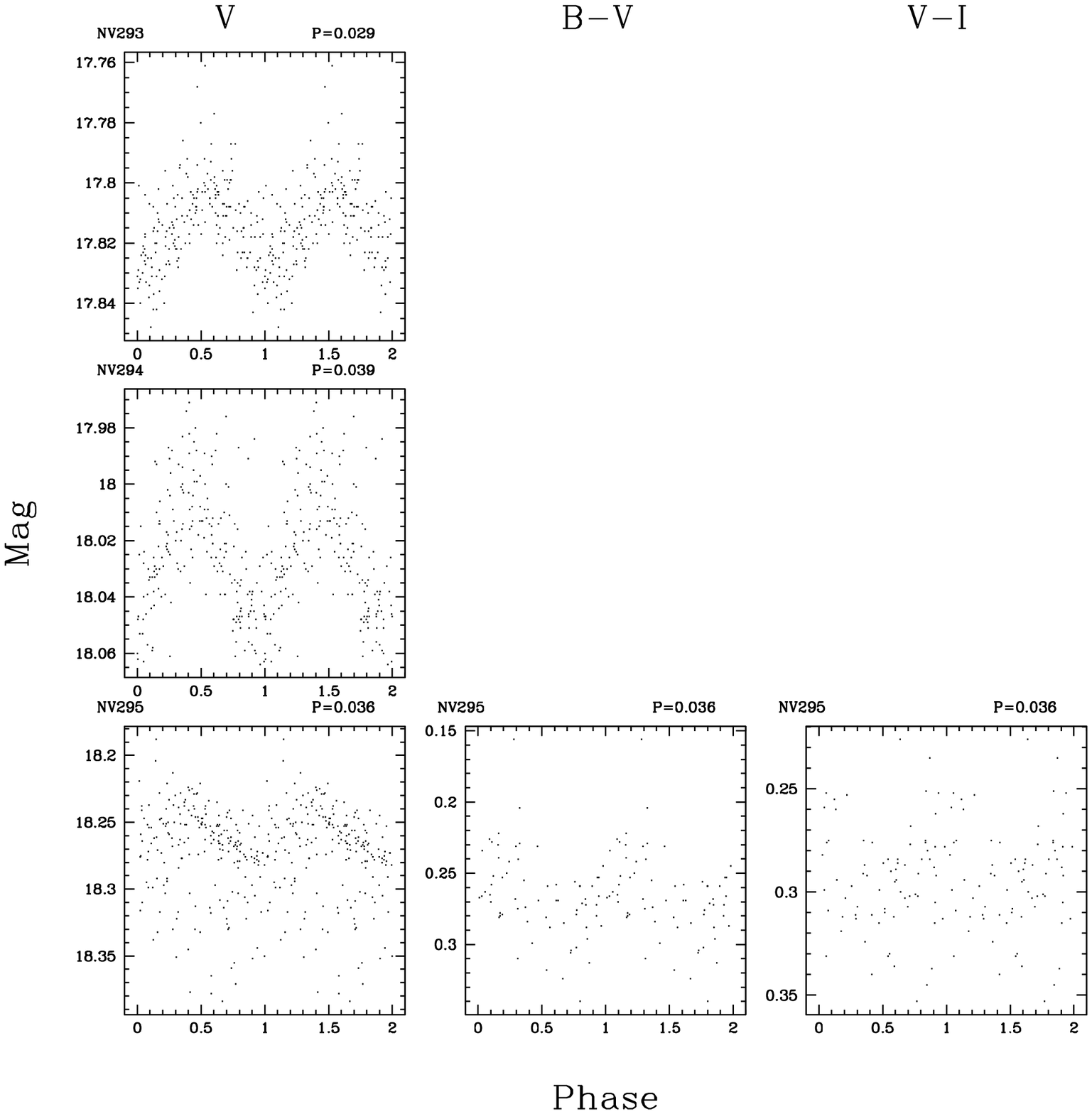}
\caption{{\it Continued}.}
\end{figure}

\clearpage

\begin{figure}[p]
\epsscale{1}
\plotone{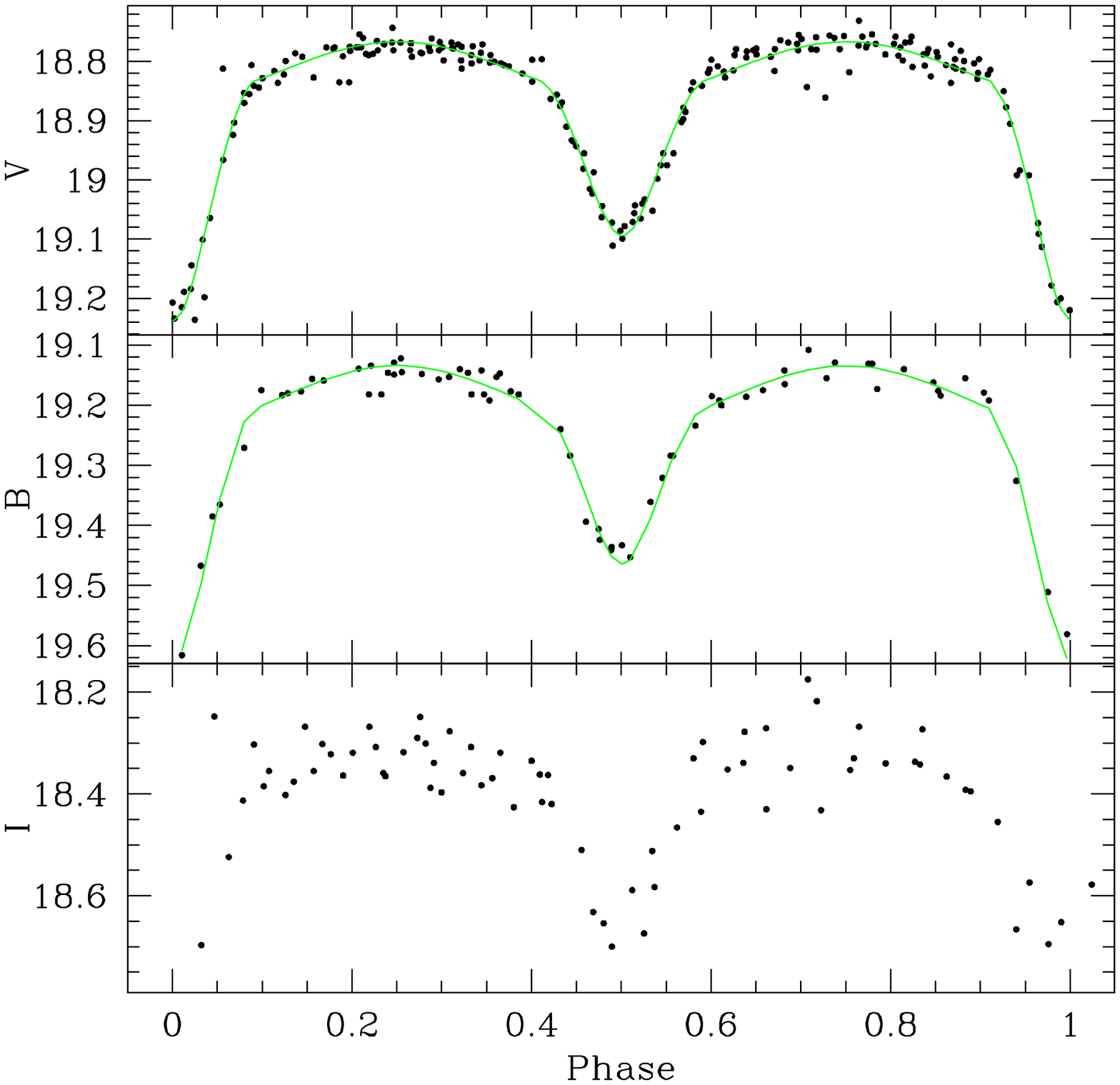}
\caption{$V$, $B$ and $I$-band light curves for the newly discovered eclipsing binary NV296. The solid line shows the model fit to the light curves using EBOP.}
\label{EB}
\end{figure}

\clearpage

\begin{figure}[p]
\epsscale{1}
\plotone{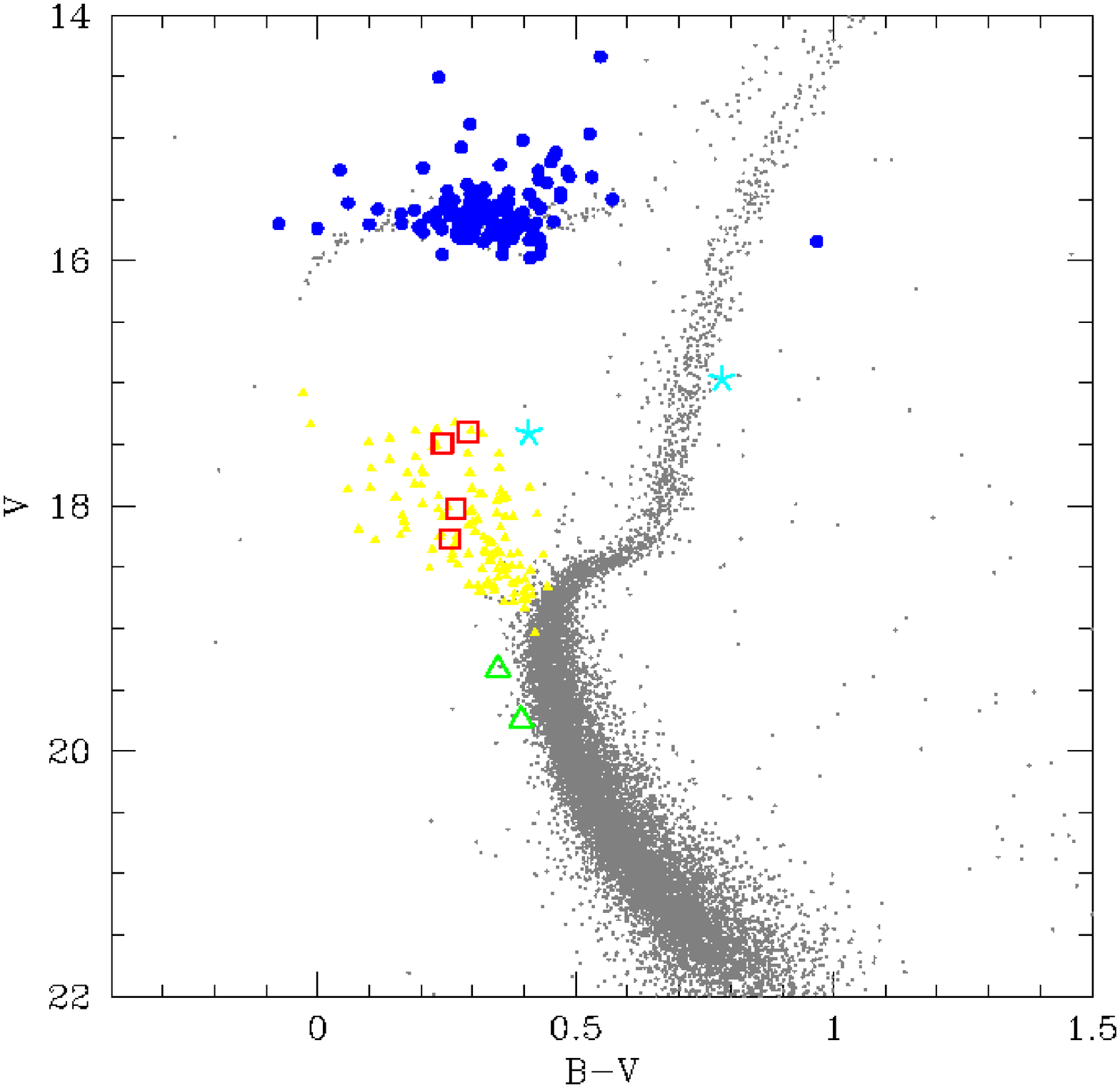}
\caption{$V$ vs $B-V$ CMD for M3. We use filled circles to show RR Lyr variables, open squares for SXP variables, stars for unclassified variables and open triangles for the two components of the EB. Filled triangles show the BSS stars that we examined for variability.}
\label{cmbv}
\end{figure}

\clearpage

\begin{figure}[p]
\epsscale{1}
\plotone{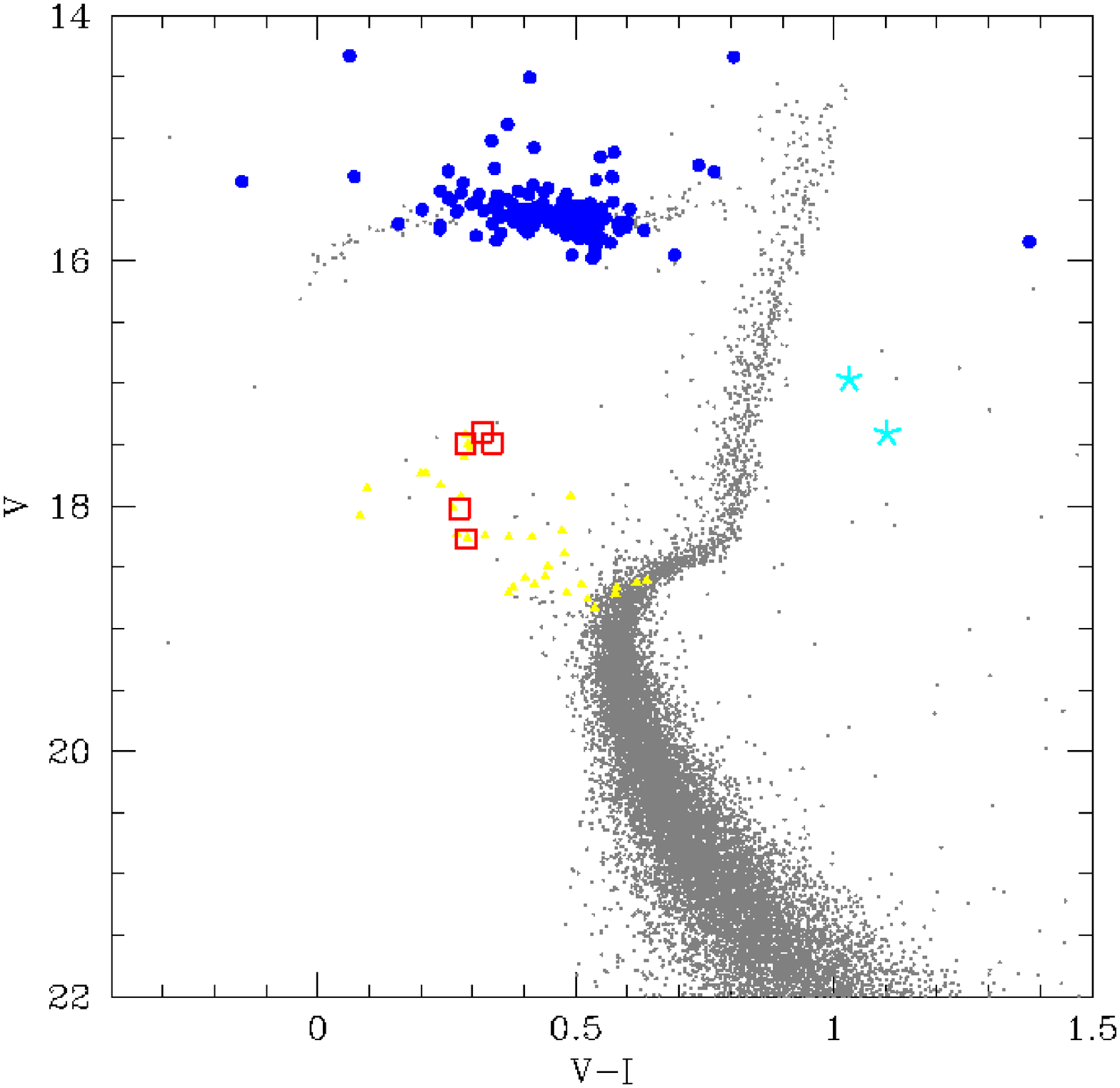}
\caption{$V$ vs $V-I$ CMD for M3. Symbols are as in Figure~\ref{cmbv}. The eclipsing binary is not shown on this diagram as we did not obtain a fit to the I-band light curve.}
\label{cmvi}
\end{figure}

\clearpage

\begin{figure}[p]
\epsscale{1}
\plotone{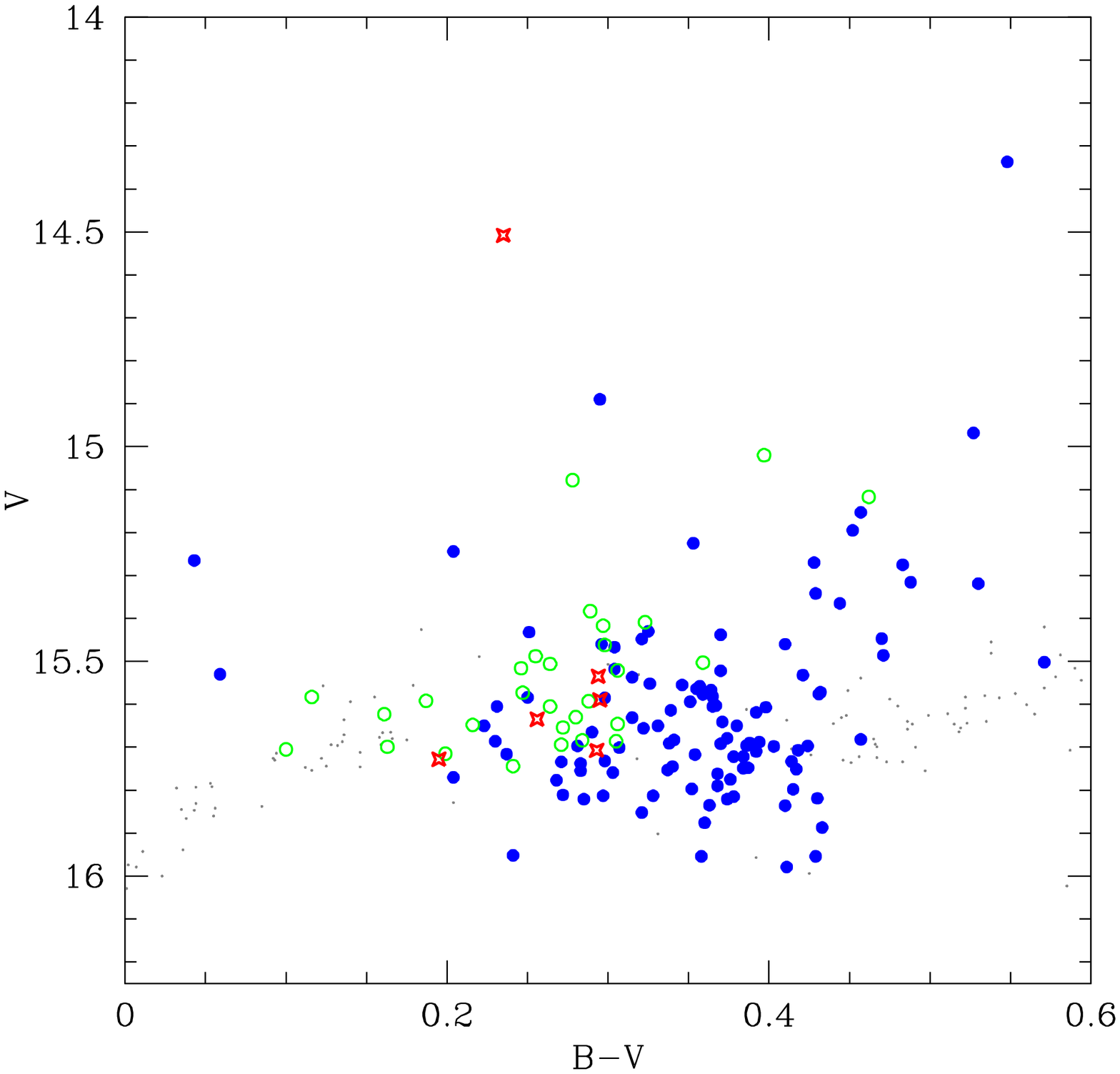}
\caption{$V$ vs $B-V$ CMD for the horizontal branch of M3. Here we distinguish between the types of RR Lyr variables using filled circles for RR0, open circles for RR1 and open stars for RR01 candidates.}
\label{cmbv_hb}
\end{figure}

\clearpage

\begin{figure}[p]
\epsscale{1}
\plotone{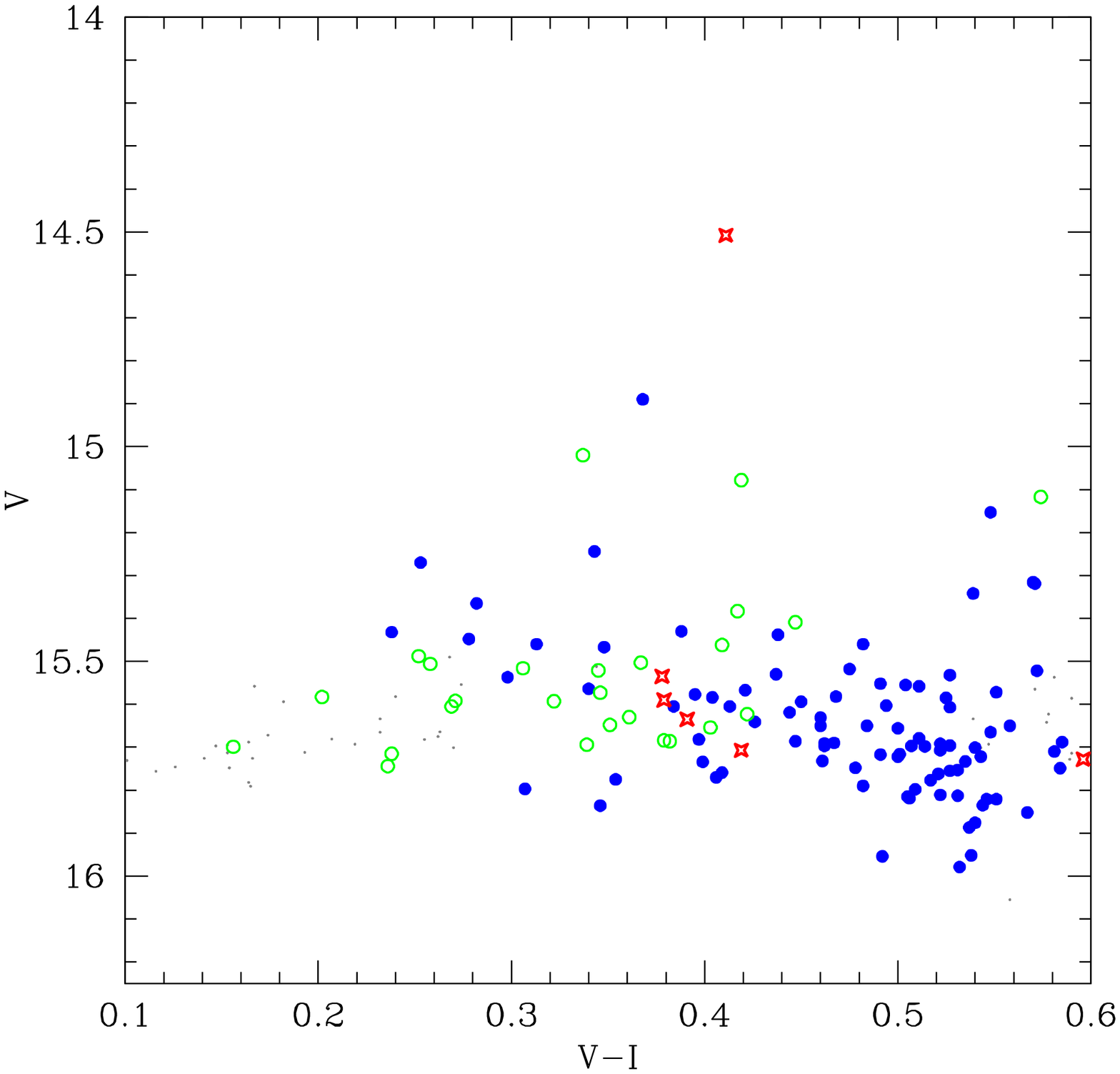}
\caption{$V$ vs $V-I$ CMD for the horizontal branch of M3. Symbols are as in Figure~\ref{cmbv_hb}.}
\label{cmvi_hb}
\end{figure}

\clearpage

\begin{deluxetable}{lrr}
\tabletypesize{\footnotesize}
\tablewidth{0pc}
\tablecaption{Median Color of M3 Turnoff Stars}
\tablehead{\colhead{Chip} & \colhead{B-V} & \colhead{V-I}}
\startdata
1 & 0.450 & 0.592 \\ 2 & 0.456 & 0.580 \\ 3 & 0.450 & 0.561 \\ 4 &
0.473 & 0.590 \\
\enddata
\label{tab:tab1}
\end{deluxetable}

\begin{deluxetable}{lrrrr}
\tabletypesize{\footnotesize}
\tablewidth{0pc}
\tablecaption{EBOP Parameters for the EB NV296. Note that parameters for which the error listed is identical to zero were held fixed in the fitting process. All other errors are taken from the formal errors given by EBOP. The first two columns display results for a mass ratio of $1.0$ the last two display the results for a mass ratio of $0.7$. $J_{s}$ and $J_{p}$ is the surface brightness of the secondary and primary respectively, $R_{s}$ is the radius of the secondary, and $a$ is the semi-major axis.}
\tablehead{\colhead{Parameter} & \colhead{Value (q=1.0)} & \colhead{Standard Error} & \colhead{Value (q=0.7)} & \colhead{Standard Error}}
\startdata
Mass Ratio & 1.0 & 0.0 & 0.7 & 0.0 \\
V Primary & 19.331 & 0.028 & 19.351 & 0.024 \\
V Secondary & 19.747 & 0.042 & 19.719 & 0.036 \\
(B-V) Primary & 0.349 & 0.042 & 0.348 & 0.036 \\
(B-V) Secondary & 0.395 & 0.064 & 0.396 & 0.055 \\
V $J_{s}/J_{p}$ & 0.65922 & 0.02325 & 0.64884 & 0.02280 \\
B $J_{s}/J_{p}$ & 0.63197 & 0.03088 & 0.62028 & 0.02917 \\
$R_{p}/a$ & 0.28594 & 0.00357 & 0.28499 & 0.00360 \\
$R_{s}/R_{p}$ & 1.01456 & 0.02725 & 1.00764 & 0.02126 \\
V Limb Darkening & 0.57 & 0.00 & 0.57 & 0.00 \\
B Limb Darkening & 0.72 & 0.00 & 0.72 & 0.00 \\
Inclination & $75.66777\arcdeg$ & $0.50846\arcdeg$ & $75.86569\arcdeg$ & $0.47571\arcdeg$ \\
Eccentricity & 0.00 & 0.00 & 0.00 & 0.00 \\
V Scale Factor & 18.76666 & 0.00214 & 18.76719 & 0.00228 \\
B Scale Factor & 19.13395 & 0.00230 & 19.13484 & 0.00237 \\
\enddata
\label{Tab:EB}
\end{deluxetable}

\begin{deluxetable}{lcrrcccrrrrrr}
\tabletypesize{\footnotesize}
\tablewidth{0pc}
\tablecaption{Catalog of M3 Variables: First 12 Columns}
\tablehead{\colhead{ID} & \colhead{{\it 2MASS}} & \colhead{$\alpha_{2000}$} &
\colhead{$\delta_{2000}$} &\colhead{$V_{flag}$} & \colhead{$B_{flag}$} &
\colhead{$I_{flag}$} &\colhead{$A_{V}$} & \colhead{$\langle V\rangle$} &
\colhead{$A_{B}$} & \colhead{$\langle B\rangle$} & \colhead{$A_{I}$} &
\colhead{$\langle I\rangle$}}
\startdata
V001 & 1 & 13:42:11.12 & +28:20:33.8 & 1 & 1 & 1 & 01.090 & 15.722 & 01.381 & 16.114 & 00.645 & 15.226 \\
V004 & 1 & 13:42:08.21 & +28:22:33.3 & 1 & 0 & 1 & 00.784 & 14.304 & \nodata & \nodata & 00.441 & 14.259 \\
V005 & 1 & 13:42:31.29 & +28:22:20.7 & 1 & 1 & 1 & 00.848 & 15.780 & 01.045 & 16.143 & 00.504 & 15.295 \\
V006 & 1 & 13:42:02.08 & +28:23:41.6 & 1 & 1 & 1 & 01.108 & 15.782 & 01.362 & 16.155 & 00.687 & 15.301 \\
V007 & 0 & 13:42:11.09 & +28:24:10.6 & 1 & 1 & 1 & 01.241 & 15.515 & 01.470 & 15.818 & 00.755 & 15.229 \\
V009 & 1 & 13:41:49.51 & +28:19:13.3 & 1 & 1 & 1 & 01.058 & 15.686 & 01.300 & 16.017 & 00.647 & 15.216 \\
V010 & 1 & 13:42:23.10 & +28:25:00.6 & 1 & 1 & 1 & 00.948 & 15.613 & 01.097 & 15.919 & 00.594 & 15.167 \\
V011 & 1 & 13:41:59.98 & +28:19:11.8 & 1 & 1 & 1 & 01.214 & 15.753 & 01.503 & 16.174 & 00.767 & 15.271 \\
V012 & 1 & 13:42:11.26 & +28:20:17.0 & 1 & 1 & 1 & 00.540 & 15.591 & 00.632 & 15.876 & 00.294 & 15.272 \\
V017 & 1 & 13:42:22.42 & +28:15:22.7 & 1 & 1 & 1 & 00.824 & 15.684 & 00.972 & 15.972 & 00.474 & 15.188 \\
V018 & 1 & 13:42:18.95 & +28:17:47.3 & 1 & 1 & 1 & 01.151 & 15.703 & 01.306 & 15.979 & 00.673 & 15.263 \\
V019 & 1 & 13:42:38.12 & +28:18:37.8 & 1 & 1 & 1 & 00.507 & 15.745 & 00.589 & 16.119 & 00.290 & 15.162 \\
V020 & 1 & 13:42:36.83 & +28:18:11.8 & 1 & 1 & 1 & 01.025 & 15.588 & 01.209 & 15.846 & 00.583 & 15.186 \\
V021 & 1 & 13:42:37.78 & +28:23:01.4 & 1 & 1 & 1 & 01.084 & 15.778 & 01.343 & 16.070 & 00.669 & 15.271 \\
V022 & 1 & 13:42:25.92 & +28:22:32.1 & 1 & 1 & 1 & 01.220 & 15.913 & 01.427 & 16.236 & 00.801 & 15.443 \\
V023 & 1 & 13:42:02.84 & +28:27:20.9 & 1 & 1 & 1 & 00.885 & 15.642 & 01.047 & 15.932 & 00.583 & 15.112 \\
V024 & 1 & 13:42:00.31 & +28:22:51.9 & 1 & 1 & 1 & 00.713 & 15.523 & 00.885 & 15.939 & 00.431 & 15.001 \\
V027 & 1 & 13:42:03.17 & +28:20:58.9 & 1 & 1 & 1 & 00.863 & 15.588 & 01.078 & 15.974 & 00.528 & 15.107 \\
V030 & 0 & 13:42:08.73 & +28:23:40.3 & 1 & 0 & 0 & 01.051 & 15.312 & \nodata & \nodata & \nodata & \nodata \\
V031 & 1 & 13:42:13.97 & +28:23:47.4 & 1 & 1 & 1 & 01.150 & 15.537 & 01.418 & 15.877 & 00.686 & 15.135 \\
V032 & 1 & 13:42:12.39 & +28:23:42.3 & 1 & 1 & 1 & 01.098 & 15.430 & 01.279 & 15.732 & 00.701 & 15.163 \\
V034 & 1 & 13:42:21.71 & +28:25:32.5 & 1 & 1 & 1 & 00.488 & 15.682 & 00.622 & 16.060 & 00.274 & 15.221 \\
V036 & 1 & 13:42:24.55 & +28:22:07.4 & 1 & 1 & 1 & 01.146 & 15.581 & 01.308 & 15.926 & 00.738 & 15.211 \\
V037 & 1 & 13:41:53.56 & +28:25:25.6 & 1 & 1 & 1 & 00.493 & 15.679 & 00.614 & 15.958 & 00.300 & 15.304 \\
V038 & 1 & 13:41:56.04 & +28:24:49.1 & 1 & 1 & 1 & 01.002 & 15.601 & 01.292 & 15.975 & 00.618 & 15.168 \\
V039 & 1 & 13:41:52.97 & +28:24:42.4 & 1 & 1 & 1 & 00.918 & 15.677 & 01.160 & 16.070 & 00.570 & 15.177 \\
V040 & 1 & 13:41:50.93 & +28:24:33.1 & 1 & 1 & 1 & 00.960 & 15.727 & 01.158 & 16.060 & 00.602 & 15.216 \\
V043 & 1 & 13:42:19.07 & +28:23:07.0 & 1 & 1 & 1 & 01.112 & 15.625 & 01.265 & 15.847 & 00.672 & 15.188 \\
V044 & 1 & 13:42:24.37 & +28:24:22.0 & 1 & 1 & 1 & 01.201 & 15.623 & 01.240 & 15.986 & 00.684 & 15.207 \\
V045 & 1 & 13:41:53.23 & +28:20:31.1 & 1 & 1 & 1 & 00.899 & 15.786 & 01.164 & 16.093 & 02.607 & 13.893 \\
V050 & 1 & 13:42:12.24 & +28:18:47.8 & 1 & 1 & 1 & 00.688 & 15.581 & 00.914 & 15.821 & 00.430 & 15.180 \\
V051 & 1 & 13:42:13.87 & +28:18:55.9 & 1 & 1 & 1 & 00.881 & 15.713 & 01.025 & 16.109 & 00.498 & 15.191 \\
V053 & 0 & 13:42:10.94 & +28:24:45.1 & 1 & 1 & 1 & 01.184 & 15.695 & 01.389 & 15.945 & 00.729 & 15.319 \\
V054 & 1 & 13:42:08.98 & +28:24:28.5 & 1 & 1 & 1 & 01.029 & 15.751 & 01.216 & 16.094 & 00.579 & 15.295 \\
V055 & 1 & 13:41:55.91 & +28:28:05.7 & 1 & 1 & 1 & 01.059 & 15.809 & 01.350 & 16.120 & 00.663 & 15.274 \\
V056 & 1 & 13:42:00.67 & +28:28:39.9 & 1 & 1 & 1 & 00.488 & 15.629 & 00.613 & 15.904 & 00.299 & 15.269 \\
V057 & 1 & 13:42:23.25 & +28:22:42.3 & 1 & 1 & 1 & 00.682 & 15.875 & 00.932 & 16.292 & 00.440 & 15.347 \\
V059 & 1 & 13:42:03.25 & +28:18:53.2 & 1 & 1 & 1 & 00.889 & 15.686 & 00.979 & 16.107 & 00.473 & 15.182 \\
V060 & 1 & 13:41:49.05 & +28:17:25.5 & 1 & 1 & 1 & 00.699 & 15.561 & 00.882 & 15.974 & 00.446 & 15.023 \\
V061 & 1 & 13:42:25.78 & +28:28:45.7 & 1 & 1 & 1 & 00.705 & 15.732 & 00.885 & 16.103 & 00.403 & 15.262 \\
V062 & 1 & 13:42:18.22 & +28:29:39.1 & 1 & 1 & 1 & 00.534 & 15.645 & 00.607 & 16.025 & 00.346 & 15.093 \\
V063 & 1 & 13:42:14.21 & +28:28:24.1 & 1 & 1 & 1 & 00.721 & 15.711 & 00.868 & 16.082 & 00.428 & 15.221 \\
V064 & 1 & 13:42:20.11 & +28:28:12.5 & 1 & 1 & 1 & 00.719 & 15.689 & 00.845 & 16.058 & 00.441 & 15.169 \\
V065 & 1 & 13:42:20.92 & +28:28:10.0 & 1 & 1 & 1 & 00.933 & 15.566 & 01.109 & 15.852 & 00.591 & 15.056 \\
V066 & 1 & 13:42:03.77 & +28:24:42.8 & 1 & 0 & 0 & 00.542 & 15.680 & \nodata & \nodata & \nodata & \nodata \\
V067 & 1 & 13:42:01.51 & +28:24:44.3 & 1 & 1 & 1 & 01.019 & 15.680 & 01.249 & 15.981 & 00.638 & 15.156 \\
V068 & 1 & 13:42:13.09 & +28:25:37.1 & 1 & 1 & 1 & 00.803 & 15.619 & 00.961 & 15.880 & 00.465 & 15.238 \\
V069 & 1 & 13:42:17.57 & +28:25:03.3 & 1 & 1 & 1 & 00.929 & 15.679 & 01.062 & 16.083 & 00.568 & 15.229 \\
V070 & 1 & 13:42:14.32 & +28:25:14.4 & 1 & 1 & 1 & 00.394 & 15.404 & 00.473 & 15.723 & 00.246 & 14.961 \\
V071 & 1 & 13:42:23.65 & +28:22:40.5 & 1 & 1 & 1 & 00.970 & 15.692 & 01.188 & 15.936 & 00.591 & 15.210 \\
V072 & 1 & 13:42:45.25 & +28:22:41.3 & 1 & 1 & 1 & 01.270 & 15.733 & 01.521 & 15.912 & 00.793 & 15.351 \\
V073 & 1 & 13:42:44.72 & +28:23:45.7 & 1 & 1 & 1 & 00.266 & 15.707 & 00.352 & 16.097 & 00.181 & 15.130 \\
V074 & 1 & 13:42:18.15 & +28:25:13.1 & 1 & 1 & 1 & 01.125 & 15.950 & 01.489 & 16.329 & 00.583 & 15.435 \\
V075 & 1 & 13:42:15.13 & +28:25:21.4 & 1 & 1 & 1 & 00.541 & 15.652 & 00.610 & 15.867 & 00.314 & 15.299 \\
V078 & 1 & 13:42:15.06 & +28:23:48.5 & 1 & 1 & 1 & 00.872 & 15.539 & 01.064 & 15.884 & 00.510 & 15.043 \\
V079 & 1 & 13:42:14.69 & +28:28:31.4 & 1 & 1 & 1 & 00.559 & 15.703 & 00.565 & 16.000 & 00.310 & 15.287 \\
V080 & 1 & 13:42:43.02 & +28:27:27.8 & 1 & 1 & 1 & 01.020 & 15.790 & 01.240 & 16.141 & 00.639 & 15.273 \\
V081 & 1 & 13:42:37.36 & +28:28:34.3 & 1 & 1 & 1 & 01.069 & 15.740 & 01.234 & 16.007 & 00.657 & 15.248 \\
V083 & 1 & 13:41:38.00 & +28:24:33.4 & 1 & 1 & 1 & 01.166 & 15.822 & 01.396 & 16.170 & 00.764 & 15.318 \\
V084 & 1 & 13:42:16.31 & +28:25:27.1 & 1 & 1 & 1 & 00.734 & 15.681 & 00.844 & 16.043 & 00.452 & 15.168 \\
V085 & 1 & 13:42:34.65 & +28:26:29.0 & 1 & 1 & 1 & 00.506 & 15.516 & 00.613 & 15.757 & 00.316 & 15.209 \\
V087 & 1 & 13:42:19.84 & +28:23:42.6 & 1 & 1 & 1 & 00.571 & 15.535 & 00.631 & 15.826 & 00.345 & 15.157 \\
V090 & 1 & 13:42:18.92 & +28:19:34.2 & 1 & 1 & 1 & 01.096 & 15.658 & 01.335 & 15.971 & 00.648 & 15.219 \\
V091 & 0 & 13:42:10.59 & +28:13:32.2 & 2 & 2 & 2 & \nodata & \nodata & \nodata & \nodata & \nodata & \nodata \\
V093 & 1 & 13:41:47.40 & +28:16:04.1 & 1 & 1 & 1 & 02.461 & 15.768 & 00.947 & 16.013 & 02.472 & 15.252 \\
V094 & 1 & 13:41:34.54 & +28:18:55.1 & 1 & 1 & 1 & 01.141 & 15.775 & 01.342 & 16.107 & 00.681 & 15.256 \\
V096 & 1 & 13:41:59.12 & +28:18:47.3 & 1 & 1 & 1 & 00.786 & 15.941 & 00.451 & 16.370 & 00.574 & 15.254 \\
V097 & 1 & 13:42:01.70 & +28:19:24.6 & 1 & 1 & 1 & 00.445 & 15.685 & 00.550 & 15.995 & 00.262 & 15.305 \\
V099 & 1 & 13:42:26.76 & +28:21:48.0 & 1 & 1 & 1 & 00.549 & 15.588 & 00.700 & 15.873 & 00.334 & 15.210 \\
V100 & 1 & 13:42:16.75 & +28:24:19.7 & 1 & 1 & 1 & 00.616 & 15.713 & 00.721 & 16.086 & 00.377 & 15.178 \\
V101 & 1 & 13:42:14.98 & +28:24:05.6 & 1 & 1 & 1 & 00.627 & 15.671 & 00.714 & 16.036 & 00.401 & 15.167 \\
V104 & 1 & 13:42:09.49 & +28:25:07.3 & 1 & 1 & 1 & 01.170 & 15.495 & 01.365 & 15.789 & 00.771 & 15.039 \\
V105 & 1 & 13:42:09.85 & +28:25:53.2 & 1 & 1 & 1 & 00.336 & 15.591 & 00.417 & 15.776 & 00.193 & 15.320 \\
V108 & 1 & 13:41:54.77 & +28:27:51.9 & 1 & 1 & 1 & 01.122 & 15.661 & 01.398 & 15.892 & 00.716 & 15.238 \\
V117 & 1 & 13:42:18.39 & +28:14:54.0 & 1 & 1 & 1 & 00.674 & 15.647 & 00.852 & 15.988 & 00.383 & 15.156 \\
V118 & 1 & 13:42:22.50 & +28:17:50.6 & 1 & 1 & 1 & 01.205 & 15.886 & 01.444 & 16.204 & 00.730 & 15.391 \\
V119 & 1 & 13:42:30.67 & +28:24:28.8 & 1 & 1 & 1 & 01.121 & 15.452 & 01.387 & 15.736 & 00.712 & 15.144 \\
V120 & 1 & 13:41:48.99 & +28:26:32.2 & 1 & 1 & 1 & 00.437 & 15.683 & 00.545 & 16.076 & 00.277 & 15.103 \\
V121 & 0 & 13:42:08.17 & +28:23:38.1 & 1 & 0 & 1 & 01.165 & 15.327 & \nodata & \nodata & 00.906 & 15.484 \\
V122 & 0 & 13:42:09.03 & +28:21:55.7 & 2 & 2 & 2 & \nodata & \nodata & \nodata & \nodata & \nodata & \nodata \\
V125 & 1 & 13:42:25.64 & +28:20:30.1 & 1 & 1 & 1 & 00.458 & 15.654 & 00.534 & 15.917 & 00.260 & 15.253 \\
V128 & 0 & 13:42:20.12 & +28:24:53.9 & 1 & 1 & 1 & 00.546 & 15.605 & 00.645 & 15.856 & 00.319 & 15.334 \\
V129 & 0 & 13:42:08.21 & +28:23:59.7 & 1 & 0 & 0 & 00.439 & 15.034 & \nodata & \nodata & \nodata & \nodata \\
V130 & 1 & 13:42:11.77 & +28:24:06.2 & 1 & 1 & 1 & 00.439 & 15.548 & 00.515 & 15.876 & 00.266 & 15.060 \\
V134 & 0 & 13:42:09.78 & +28:23:35.0 & 1 & 1 & 1 & 00.642 & 15.644 & 00.725 & 15.975 & 00.402 & 15.166 \\
V135 & 1 & 13:42:09.43 & +28:23:20.4 & 1 & 1 & 1 & 00.820 & 15.539 & 01.096 & 15.885 & 00.490 & 15.052 \\
V136 & 0 & 13:42:09.57 & +28:23:16.5 & 1 & 1 & 1 & 00.558 & 15.311 & 00.656 & 15.797 & 00.303 & 14.746 \\
V137 & 1 & 13:42:15.47 & +28:22:23.2 & 1 & 1 & 1 & 00.898 & 15.572 & 00.968 & 15.910 & 00.578 & 15.139 \\
V139 & 0 & 13:42:14.11 & +28:23:10.8 & 1 & 1 & 1 & 01.240 & 15.531 & 01.376 & 15.865 & 00.803 & 15.211 \\
V140 & 1 & 13:42:10.28 & +28:24:30.7 & 1 & 1 & 1 & 00.507 & 15.515 & 00.618 & 15.815 & 00.259 & 15.176 \\
V142 & 1 & 13:42:09.26 & +28:21:43.6 & 1 & 1 & 1 & 01.075 & 15.775 & 01.212 & 16.115 & 00.772 & 15.483 \\
V143 & 1 & 13:42:08.88 & +28:22:59.0 & 1 & 1 & 1 & 01.073 & 15.416 & 01.321 & 15.769 & 00.598 & 14.989 \\
V145 & 0 & 13:42:13.61 & +28:22:51.3 & 1 & 1 & 1 & 00.943 & 15.429 & 01.207 & 15.820 & 00.566 & 14.968 \\
V146 & 0 & 13:42:18.48 & +28:21:44.3 & 1 & 1 & 1 & 01.090 & 15.649 & 01.410 & 16.067 & 00.737 & 15.273 \\
V147 & 0 & 13:42:09.83 & +28:23:29.7 & 1 & 1 & 1 & 00.474 & 15.693 & 00.475 & 15.963 & 00.265 & 15.354 \\
V148 & 1 & 13:42:10.95 & +28:23:20.0 & 1 & 1 & 1 & 00.832 & 15.194 & 01.108 & 15.558 & 00.404 & 14.479 \\
V149 & 0 & 13:42:14.10 & +28:23:35.5 & 1 & 1 & 1 & 01.148 & 15.642 & 01.289 & 15.966 & 00.594 & 15.070 \\
V150 & 0 & 13:42:16.68 & +28:23:20.8 & 1 & 1 & 1 & 00.827 & 15.798 & 00.956 & 16.166 & 00.499 & 15.268 \\
V151 & 0 & 13:42:11.99 & +28:22:01.3 & 1 & 1 & 1 & 00.909 & 15.459 & 01.038 & 15.757 & 00.566 & 15.116 \\
V152 & 1 & 13:42:17.47 & +28:23:33.6 & 1 & 1 & 1 & 00.458 & 15.490 & 00.509 & 15.733 & 00.272 & 15.235 \\
V156 & 0 & 13:42:09.99 & +28:22:01.8 & 1 & 1 & 1 & 01.005 & 15.415 & 01.163 & 15.726 & 00.511 & 15.039 \\
V157 & 1 & 13:42:10.20 & +28:23:18.8 & 1 & 1 & 1 & 01.078 & 15.249 & 01.261 & 15.707 & 00.412 & 14.505 \\
V159 & 0 & 13:42:10.34 & +28:22:59.1 & 1 & 1 & 1 & 00.432 & 15.312 & 00.586 & 15.841 & 00.388 & 14.747 \\
V160 & 0 & 13:42:10.80 & +28:21:59.4 & 1 & 1 & 1 & 01.046 & 15.567 & 01.205 & 15.932 & 00.668 & 15.115 \\
V161 & 0 & 13:42:12.80 & +28:21:45.2 & 1 & 1 & 1 & 00.681 & 15.730 & 00.895 & 16.132 & 00.368 & 15.112 \\
V165 & 0 & 13:42:17.04 & +28:23:03.2 & 1 & 1 & 2 & 01.076 & 15.468 & 01.401 & 16.006 & \nodata & \nodata \\
V168 & 1 & 13:42:08.09 & +28:22:49.9 & 1 & 0 & 1 & 00.473 & 15.192 & \nodata & \nodata & 00.462 & 15.595 \\
V170 & 1 & 13:42:09.33 & +28:23:15.2 & 1 & 1 & 1 & 00.466 & 15.246 & 00.595 & 15.453 & 00.271 & 14.901 \\
V171 & 0 & 13:42:09.51 & +28:22:59.6 & 1 & 1 & 1 & 00.567 & 15.580 & 00.618 & 15.703 & 00.380 & 15.381 \\
V172 & 0 & 13:42:09.89 & +28:23:08.8 & 1 & 1 & 0 & 01.021 & 15.590 & 01.108 & 15.927 & \nodata & \nodata \\
V173 & 0 & 13:42:10.50 & +28:23:21.8 & 1 & 1 & 1 & 00.526 & 14.881 & 00.563 & 15.170 & 00.294 & 14.520 \\
V174 & 0 & 13:42:10.84 & +28:22:08.9 & 1 & 1 & 1 & 01.174 & 15.735 & 01.343 & 16.080 & 00.761 & 15.404 \\
V175 & 0 & 13:42:14.62 & +28:23:09.4 & 1 & 1 & 0 & 00.932 & 15.713 & 00.844 & 15.714 & \nodata & \nodata \\
V176 & 0 & 13:42:14.98 & +28:23:16.2 & 1 & 1 & 0 & 01.004 & 15.717 & 01.141 & 15.989 & \nodata & \nodata \\
V177 & 1 & 13:42:16.28 & +28:22:13.9 & 1 & 1 & 1 & 00.557 & 15.507 & 00.674 & 15.765 & 00.330 & 15.250 \\
V178 & 0 & 13:42:17.47 & +28:23:29.9 & 1 & 1 & 1 & 00.406 & 15.712 & 00.448 & 15.908 & 00.245 & 15.475 \\
V180 & 0 & 13:42:10.09 & +28:22:13.5 & 1 & 1 & 0 & 00.770 & 15.735 & 00.832 & 16.071 & \nodata & \nodata \\
V181 & 0 & 13:42:09.21 & +28:22:29.3 & 1 & 1 & 0 & 00.410 & 15.188 & 00.475 & 15.636 & \nodata & \nodata \\
V184 & 0 & 13:42:09.57 & +28:22:27.8 & 1 & 1 & 1 & 01.337 & 15.770 & 01.467 & 16.039 & 01.045 & 15.277 \\
V187 & 0 & 13:42:09.64 & +28:22:52.0 & 2 & 2 & 2 & \nodata & \nodata & \nodata & \nodata & \nodata & \nodata \\
V188 & 0 & 13:42:09.49 & +28:23:06.9 & 1 & 1 & 1 & 00.449 & 15.742 & 00.477 & 15.982 & 00.337 & 15.506 \\
V189 & 0 & 13:42:09.59 & +28:22:22.0 & 1 & 1 & 1 & 00.478 & 15.146 & 00.629 & 15.609 & 00.304 & 14.603 \\
V190 & 0 & 13:42:10.88 & +28:23:11.4 & 1 & 1 & 1 & 01.095 & 15.557 & 01.276 & 15.905 & 00.713 & 15.177 \\
V191 & 0 & 13:42:11.60 & +28:23:06.0 & 2 & 2 & 2 & \nodata & \nodata & \nodata & \nodata & \nodata & \nodata \\
V192 & 0 & 13:42:11.31 & +28:22:46.9 & 1 & 1 & 1 & 00.308 & 14.328 & 00.454 & 14.868 & 00.153 & 13.523 \\
V193 & 0 & 13:42:12.61 & +28:22:35.8 & 1 & 1 & 1 & 00.811 & 15.255 & 01.000 & 15.674 & 00.615 & 15.010 \\
V194 & 0 & 13:42:12.74 & +28:22:30.2 & 1 & 1 & 2 & 00.621 & 15.250 & 00.476 & 15.294 & \nodata & \nodata \\
V195 & 0 & 13:42:10.48 & +28:22:14.8 & 1 & 1 & 0 & 00.389 & 15.444 & 00.376 & 15.914 & \nodata & \nodata \\
V197 & 0 & 13:42:15.91 & +28:22:52.2 & 1 & 1 & 1 & 01.347 & 15.505 & 01.455 & 15.586 & 00.764 & 15.089 \\
V200 & 0 & 13:42:11.19 & +28:23:04.1 & 2 & 2 & 0 & \nodata & \nodata & \nodata & \nodata & \nodata & \nodata \\
V201 & 0 & 13:42:11.78 & +28:22:34.0 & 1 & 1 & 1 & 00.847 & 15.345 & 01.058 & 15.778 & 00.554 & 15.079 \\
V202 & 1 & 13:41:42.83 & +28:24:17.5 & 1 & 1 & 1 & 00.142 & 15.574 & 00.214 & 16.007 & 00.099 & 14.971 \\
V207 & 1 & 13:42:14.19 & +28:22:11.9 & 1 & 1 & 1 & 00.366 & 15.381 & 00.397 & 15.675 & 00.196 & 14.966 \\
V208 & 0 & 13:42:11.72 & +28:21:44.5 & 1 & 1 & 1 & 00.397 & 15.502 & 00.522 & 15.860 & 00.256 & 15.135 \\
V212 & 0 & 13:42:09.87 & +28:22:04.4 & 1 & 1 & 0 & 00.858 & 15.475 & 00.972 & 15.938 & \nodata & \nodata \\
V213 & 0 & 13:42:09.57 & +28:22:12.8 & 1 & 1 & 1 & 00.468 & 15.461 & 00.529 & 15.753 & 00.396 & 15.046 \\
V214 & 0 & 13:42:13.94 & +28:22:48.9 & 1 & 2 & 1 & 00.920 & 15.292 & \nodata & \nodata & 00.549 & 15.237 \\
V215 & 0 & 13:42:10.43 & +28:22:41.7 & 1 & 1 & 1 & 00.793 & 15.423 & 00.797 & 15.664 & 00.501 & 15.188 \\
V216 & 0 & 13:42:13.60 & +28:22:31.6 & 1 & 1 & 1 & 00.525 & 15.703 & 00.517 & 15.863 & 00.393 & 15.542 \\
V218 & 0 & 13:42:13.62 & +28:22:13.2 & 1 & 1 & 1 & 00.806 & 15.738 & 00.936 & 16.037 & 00.471 & 15.343 \\
V220 & 1 & 13:42:13.98 & +28:22:27.1 & 2 & 2 & 2 & \nodata & \nodata & \nodata & \nodata & \nodata & \nodata \\
V221 & 0 & 13:42:10.21 & +28:22:28.9 & 1 & 1 & 1 & 00.358 & 15.020 & 00.445 & 15.418 & 00.601 & 14.672 \\
V223 & 0 & 13:42:13.28 & +28:22:36.6 & 1 & 1 & 0 & 00.605 & 15.647 & 00.673 & 15.950 & \nodata & \nodata \\
V229 & 0 & 13:42:09.00 & +28:21:56.8 & 2 & 0 & 1 & \nodata & \nodata & \nodata & \nodata & 00.869 & 15.272 \\
V234 & 0 & 13:42:13.09 & +28:22:02.9 & 1 & 1 & 1 & 01.137 & 15.800 & 01.260 & 16.184 & 00.897 & 15.476 \\
V235 & 0 & 13:42:13.77 & +28:23:19.8 & 1 & 1 & 1 & 00.507 & 15.516 & 00.588 & 15.886 & 00.309 & 14.948 \\
V237 & 0 & 13:42:15.73 & +28:18:17.3 & 1 & 1 & 1 & 00.232 & 18.019 & 00.228 & 18.288 & 00.243 & 17.743 \\
V239 & 0 & 13:42:09.86 & +28:22:16.0 & 2 & 2 & 0 & \nodata & \nodata & \nodata & \nodata & \nodata & \nodata \\
V240 & 0 & 13:42:09.46 & +28:22:35.3 & 1 & 1 & 1 & 00.511 & 15.624 & 00.479 & 15.785 & 00.258 & 15.200 \\
V241 & 0 & 13:42:10.76 & +28:22:38.0 & 2 & 1 & 2 & \nodata & \nodata & 00.962 & 14.912 & \nodata & \nodata \\
V243 & 0 & 13:42:12.25 & +28:22:15.6 & 1 & 1 & 1 & 00.514 & 15.335 & 00.612 & 15.758 & 00.327 & 14.805 \\
V245 & 0 & 13:42:09.89 & +28:23:00.0 & 1 & 1 & 0 & 00.481 & 15.414 & 00.576 & 15.713 & \nodata & \nodata \\
V246 & 0 & 13:42:12.76 & +28:22:40.5 & 1 & 1 & 0 & 00.600 & 15.698 & 00.472 & 15.630 & \nodata & \nodata \\
V247 & 0 & 13:42:14.81 & +28:22:15.6 & 2 & 2 & 2 & \nodata & \nodata & \nodata & \nodata & \nodata & \nodata \\
V249 & 0 & 13:42:10.31 & +28:22:48.2 & 2 & 2 & 0 & \nodata & \nodata & \nodata & \nodata & \nodata & \nodata \\
V250 & 0 & 13:42:10.51 & +28:22:52.5 & 1 & 1 & 0 & 00.823 & 14.958 & 00.986 & 15.468 & \nodata & \nodata \\
V252 & 0 & 13:42:10.99 & +28:22:44.0 & 1 & 1 & 1 & 00.878 & 15.728 & 00.785 & 15.905 & 00.512 & 15.128 \\
V253 & 0 & 13:42:12.14 & +28:22:32.8 & 1 & 1 & 1 & 00.531 & 15.562 & 00.548 & 15.805 & 00.397 & 15.225 \\
V254 & 0 & 13:42:12.42 & +28:22:53.6 & 2 & 2 & 0 & \nodata & \nodata & \nodata & \nodata & \nodata & \nodata \\
V255 & 0 & 13:42:12.62 & +28:22:43.5 & 2 & 2 & 0 & \nodata & \nodata & \nodata & \nodata & \nodata & \nodata \\
V256 & 1 & 13:42:13.08 & +28:22:59.0 & 2 & 2 & 2 & \nodata & \nodata & \nodata & \nodata & \nodata & \nodata \\
V258 & 0 & 13:42:14.31 & +28:23:31.5 & 1 & 1 & 1 & 00.552 & 15.600 & 00.733 & 16.004 & 00.390 & 15.081 \\
V259 & 0 & 13:42:14.57 & +28:22:54.9 & 1 & 1 & 1 & 00.293 & 15.114 & 00.397 & 15.571 & 00.139 & 14.541 \\
V261 & 0 & 13:42:10.08 & +28:22:40.4 & 1 & 1 & 1 & 00.361 & 15.077 & 00.409 & 15.355 & 00.220 & 14.659 \\
V264 & 0 & 13:42:10.87 & +28:22:29.8 & 2 & 1 & 0 & \nodata & \nodata & 00.288 & 14.493 & \nodata & \nodata \\
V269 & 0 & 13:42:12.79 & +28:22:33.0 & 2 & 2 & 2 & \nodata & \nodata & \nodata & \nodata & \nodata & \nodata \\
V270 & 1 & 13:42:11.93 & +28:23:32.2 & 1 & 1 & 1 & 00.507 & 14.503 & 00.553 & 14.751 & 00.284 & 14.096 \\
V271 & 1 & 13:42:12.16 & +28:23:18.6 & 2 & 2 & 0 & \nodata & \nodata & \nodata & \nodata & \nodata & \nodata \\
NV286 & 1 & 13:41:51.55 & +28:17:00.6 & 1 & 1 & 1 & 00.171 & 17.406 & 00.104 & 17.816 & 00.128 & 16.304 \\
NV287 & 1 & 13:41:41.00 & +28:20:55.5 & 1 & 1 & 1 & 00.060 & 16.964 & 00.107 & 17.747 & 00.064 & 15.935 \\
NV288 & 0 & 13:42:17.39 & +28:13:35.1 & 1 & 1 & 1 & 00.087 & 17.489 & 00.079 & 17.731 & 00.051 & 17.204 \\
NV289 & 1 & 13:42:12.17 & +28:19:20.9 & 1 & 1 & 1 & 00.109 & 17.399 & 00.095 & 17.692 & 00.084 & 17.079 \\
NV290 & 1 & 13:42:21.30 & +28:23:45.0 & 1 & 0 & 0 & 00.060 & 15.670 & \nodata & \nodata & \nodata & \nodata \\
NV291 & 0 & 13:42:18.07 & +28:22:39.6 & 1 & 1 & 1 & 00.838 & 17.462 & 00.978 & 17.697 & 00.519 & 17.141 \\
NV292 & 0 & 13:42:11.18 & +28:21:54.0 & 1 & 1 & 0 & 00.267 & 15.704 & 00.251 & 15.809 & \nodata & \nodata \\
NV293 & 0 & 13:42:20.59 & +28:28:32.8 & 1 & 0 & 0 & 00.087 & 17.813 & \nodata & \nodata & \nodata & \nodata \\
NV294 & 0 & 13:42:13.69 & +28:25:30.3 & 1 & 0 & 0 & 00.093 & 18.024 & \nodata & \nodata & \nodata & \nodata \\
NV295 & 0 & 13:41:58.56 & +28:27:59.4 & 1 & 1 & 1 & 00.196 & 18.272 & 00.164 & 18.529 & 00.162 & 17.985 \\
NV296 & 0 & 13:41:42.98 & +28:24:04.0 & 1 & 1 & 1 & 00.466 & 18.767 & 00.310 & 19.134 & 00.398 & 18.307 \\
NV297 & 1 & 13:41:31.72 & +28:24:10.9 & 1 & 1 & 1 & 00.046 & 15.844 & 00.067 & 16.811 & 00.024 & 14.465 \\
\enddata
\tablecomments{ Full discussion of the column headings is presenting in \S3 of the text.}
\label{tab:tab2}
\end{deluxetable}

\begin{deluxetable}{lrrcrrrrl}
\tabletypesize{\footnotesize}
\tablewidth{0pc}
\tablecaption{Catalog of M3 Variables: last 9 columns}
\tablehead{\colhead{ID} & \colhead{Per} & \colhead{Per$_{pub}$} &
\colhead{Per$_{flag}$} & \colhead{$JD_{min}$} & \colhead{$B_{phase}$} & 
\colhead {$I_{phase}$} & \colhead{Class} &\colhead{Remarks}}
\startdata
V001 & 0.520300 & 0.520596 & 0 & 920.705400 & 0.268883 & 0.305977 & RR0 & Bl? \\ 
V004 & 0.585033 & 0.584900 & 0 & 920.753434 & \nodata & 0.109739 & RR0 & Bl? \\ 
V005 & 0.501451 & 0.504178 & 1 & 920.760300 & 0.107879 & 0.990228 & RR0 & Bl \\ 
V006 & 0.514333 & 0.514333 & 0 & 920.835834 & 0.672006 & 0.009527 & RR0 & \\ 
V007 & 0.497396 & 0.497429 & 0 & 920.866716 & 0.805403 & 0.974837 & RR0 & \\ 
V009 & 0.541545 & 0.541553 & 0 & 921.072365 & 0.006666 & 0.002899 & RR0 & \\ 
V010 & 0.569430 & 0.569544 & 0 & 920.748500 & 0.720984 & 0.048996 & RR0 & \\ 
V011 & 0.507910 & 0.507894 & 0 & 920.871500 & 0.999291 & 0.015948 & RR0 & \\ 
V012 & 0.317968 & 0.317540 & 1 & 920.737396 & 0.980589 & 0.070523 & RR1 & \\ 
V017 & 0.575998 & 0.576159 & 1 & 921.191002 & 0.066674 & 0.016840 & RR0 & Bl? \\ 
V018 & 0.516411 & 0.516451 & 0 & 920.928300 & 0.078831 & 0.072229 & RR0 & \\ 
V019 & 0.631976 & 0.631972 & 0 & 920.803000 & 0.001000 & 0.009766 & RR0 & Bl? \\ 
V020 & 0.491302 & 0.490476 & 1 & 920.785288 & 0.992062 & 0.130673 & RR0 & \\ 
V021 & 0.515786 & 0.515756 & 0 & 920.939100 & 0.286425 & 0.116502 & RR0 & \\ 
V022 & 0.481424 & 0.481424 & 0 & 920.986068 & 0.069577 & 0.987745 & RR0 & Bl \\ 
V023 & 0.595376 & 0.595376 & 0 & 920.942172 & 0.849769 & 0.016292 & RR0 & Bl \\ 
V024 & 0.663372 & 0.663372 & 0 & 920.808884 & 0.998667 & 0.022937 & RR0 & Bl \\ 
V027 & 0.579087 & 0.579073 & 0 & 921.214913 & 0.974076 & 0.988993 & RR0 & \\ 
V030 & 0.512092 & 0.512092 & 0 & 920.833332 & \nodata & \nodata & RR0 & \\ 
V031 & 0.580720 & 0.580720 & 0 & 921.008860 & 0.061716 & 0.013776 & RR0 & Bl? \\ 
V032 & 0.495350 & 0.495350 & 0 & 921.107500 & 0.791662 & 0.006258 & RR0 & \\ 
V034 & 0.559050 & 0.560963 & 1 & 920.860200 & 0.016725 & 0.130579 & RR0 & Bl \\ 
V036 & 0.545599 & 0.545599 & 0 & 921.000907 & 0.945748 & 0.981488 & RR0 & \\ 
V037 & 0.326639 & 0.326639 & 0 & 920.967091 & 0.056919 & 0.057786 & RR1 & \\ 
V038 & 0.558185 & 0.558011 & 0 & 921.025545 & 0.909089 & 0.891219 & RR0 & Bl? \\ 
V039 & 0.587067 & 0.587067 & 0 & 921.348533 & 0.002783 & 0.025438 & RR0 & \\ 
V040 & 0.551535 & 0.551535 & 0 & 920.761430 & 0.989919 & 0.030533 & RR0 & \\ 
V043 & 0.540532 & 0.540510 & 0 & 920.975472 & 0.053333 & 0.014800 & RR0 & \\ 
V044 & 0.506254 & 0.337850 & 0 & 920.924892 & 0.953762 & 0.037728 & RR0 & Bl \\ 
V045 & 0.536073 & 0.536073 & 0 & 920.886560 & 0.073397 & 0.672720 & RR0 & Bl \\ 
V050 & 0.512891 & 0.513170 & 1 & 921.143445 & 0.934918 & 0.994740 & RR0 & \\ 
V051 & 0.583970 & 0.583970 & 0 & 920.869800 & 0.993150 & 0.002654 & RR0 & Bl \\ 
V053 & 0.504881 & 0.504881 & 0 & 920.851000 & 0.969696 & 0.001218 & RR0 & \\ 
V054 & 0.506247 & 0.506247 & 0 & 920.804800 & 0.992703 & 0.943901 & RR0 & Bl? \\ 
V055 & 0.529822 & 0.529822 & 0 & 920.891956 & 0.994043 & 0.010203 & RR0 & \\ 
V056 & 0.329605 & 0.329600 & 0 & 920.839785 & 0.002670 & 0.065487 & RR1 & \\ 
V057 & 0.512191 & 0.512191 & 0 & 920.852656 & 0.979695 & 0.070478 & RR0 & \\ 
V059 & 0.588727 & 0.588826 & 0 & 921.113473 & 0.002257 & 0.989809 & RR0 & Bl \\ 
V060 & 0.707722 & 0.707727 & 0 & 921.270814 & 0.985164 & 0.000814 & RR0 & \\ 
V061 & 0.520941 & 0.520926 & 0 & 921.142959 & 0.958763 & 0.201435 & RR0 & Bl? \\ 
V062 & 0.652418 & 0.652418 & 0 & 921.309430 & 0.915695 & 0.075022 & RR0 & \\ 
V063 & 0.570282 & 0.570382 & 0 & 920.951790 & 0.009820 & 0.042435 & RR0 & \\ 
V064 & 0.605465 & 0.605465 & 0 & 921.317740 & 0.878928 & 0.005285 & RR0 & \\ 
V065 & 0.668349 & 0.668347 & 0 & 921.393302 & 0.987002 & 0.007501 & RR0 & \\ 
V066 & 0.620055 & 0.619100 & 1 & 921.305315 & \nodata & \nodata & RR0 & Bl \\ 
V067 & 0.568333 & 0.568333 & 0 & 921.186767 & 0.008622 & 0.996189 & RR0 & Bl \\ 
V068 & 0.358690 & 0.358690 & 0 & 920.998700 & 0.015612 & 0.077755 & RR01 & \\ 
V069 & 0.566615 & 0.566615 & 0 & 921.012215 & 0.938203 & 0.930535 & RR0 & \\ 
V070 & 0.486276 & 0.486093 & 1 & 921.023112 & 0.961314 & 0.101580 & RR1 & \\ 
V071 & 0.549053 & 0.549053 & 0 & 920.855294 & 0.865707 & 0.053314 & RR0 & Bl? \\ 
V072 & 0.456078 & 0.456078 & 0 & 920.761354 & 0.747315 & 0.986384 & RR0 & \\ 
V073 & 0.673659 & 0.670799 & 1 & 920.748500 & 0.982702 & 0.941698 & RR0 & Bl \\ 
V074 & 0.492152 & 0.492152 & 0 & 920.884772 & 0.988418 & 0.310481 & RR0 & \\ 
V075 & 0.314080 & 0.314080 & 0 & 920.860200 & 0.024134 & 0.885061 & RR1 & \\ 
V078 & 0.611965 & 0.611965 & 0 & 920.930100 & 0.936434 & 0.956615 & RR0 & Bl? \\ 
V079 & 0.361269 & 0.357269 & 1 & 921.049955 & 0.984222 & 0.990569 & RR01 & \\ 
V080 & 0.538331 & 0.537556 & 1 & 921.207445 & 0.995843 & 0.171571 & RR0 & Bl? \\ 
V081 & 0.529122 & 0.529122 & 0 & 921.025146 & 0.794478 & 0.053137 & RR0 & \\ 
V083 & 0.501264 & 0.501264 & 0 & 920.849544 & 0.728933 & 0.009289 & RR0 & \\ 
V084 & 0.595732 & 0.595732 & 0 & 920.923804 & 0.009400 & 0.990096 & RR0 & \\ 
V085 & 0.355808 & 0.355817 & 0 & 920.895076 & 0.989106 & 0.003541 & RR1 & \\ 
V087 & 0.354368 & 0.357478 & 0 & 921.064864 & 0.953427 & 0.078156 & RR01 & \\ 
V090 & 0.517031 & 0.517031 & 0 & 920.900000 & 0.036340 & 0.092138 & RR0 & \\ 
V091 & 0.529369 & 0.529369 & 0 & 920.829773 & 0.217140 & 0.168763 & RR0 & Bl? \\ 
V093 & 0.602294 & 0.602294 & 0 & 921.282644 & 0.746954 & 0.013283 & RR0 & Bl? \\ 
V094 & 0.523716 & 0.523696 & 0 & 920.813968 & 0.984251 & 0.962644 & RR0 & \\ 
V096 & 0.499416 & 0.499416 & 0 & 920.938568 & 0.061472 & 0.088704 & RR0 & \\ 
V097 & 0.334944 & 0.334933 & 0 & 920.805572 & 0.056224 & 0.055317 & RR1 & \\ 
V099 & 0.482199 & 0.361696 & 1 & 921.013804 & 0.914143 & 0.113858 & RR01 & Bl? \\ 
V100 & 0.618813 & 0.618813 & 0 & 920.813061 & 0.815114 & 0.005171 & RR0 & \\ 
V101 & 0.643886 & 0.643886 & 0 & 921.284314 & 0.967451 & 0.073370 & RR0 & Bl \\ 
V104 & 0.569946 & 0.569931 & 0 & 921.171862 & 0.989999 & 0.061220 & RR0 & \\ 
V105 & 0.287744 & 0.287744 & 0 & 921.001260 & 0.007187 & 0.058274 & RR1 & \\ 
V108 & 0.519615 & 0.519615 & 0 & 921.163810 & 0.987683 & 0.024586 & RR0 & \\ 
V117 & 0.600529 & 0.597263 & 1 & 921.078313 & 0.061159 & 0.028905 & RR0 & Bl? \\ 
V118 & 0.499391 & 0.499391 & 0 & 920.871036 & 0.014818 & 0.042051 & RR0 & \\ 
V119 & 0.517692 & 0.517692 & 0 & 921.188872 & 0.043431 & 0.994754 & RR0 & \\ 
V120 & 0.640139 & 0.640139 & 0 & 920.758900 & 0.917830 & 0.003601 & RR0 & \\ 
V121 & 0.535211 & 0.535211 & 0 & 921.211745 & \nodata & 0.091366 & RR0 & Bl \\ 
V122 & 0.509621 & 0.516090 & 0 & 920.918108 & 0.274808 & 0.078937 & RR0 & \\ 
V125 & 0.349780 & 0.349823 & 0 & 920.924420 & 0.035165 & 0.998227 & RR1 & \\ 
V128 & 0.292044 & 0.292040 & 0 & 920.748500 & 0.869759 & 0.978962 & RR1 & \\ 
V129 & 0.406102 & 0.406102 & 0 & 920.957600 & \nodata & \nodata & RR0 & \\ 
V130 & 0.565181 & 0.567840 & 1 & 921.182519 & 0.897622 & 0.930886 & RR0 & Bl \\ 
V134 & 0.618057 & 0.618057 & 0 & 921.283896 & 0.015023 & 0.996254 & RR0 & \\ 
V135 & 0.568591 & 0.568397 & 0 & 921.171075 & 0.779662 & 0.978107 & RR0 & \\ 
V136 & 0.617153 & 0.617186 & 0 & 921.176847 & 0.031759 & 0.971968 & RR0 & \\ 
V137 & 0.575161 & 0.575161 & 0 & 921.070864 & 0.893209 & 0.887390 & RR0 & \\ 
V139 & 0.559970 & 0.559970 & 0 & 920.860200 & 0.007697 & 0.957783 & RR0 & \\ 
V140 & 0.333499 & 0.333499 & 0 & 920.813726 & 0.999409 & 0.121362 & RR1 & Bl? \\ 
V142 & 0.568628 & 0.568628 & 0 & 920.790800 & 0.023045 & 0.014069 & RR0 & \\ 
V143 & 0.596739 & 0.596535 & 0 & 921.340661 & 0.034987 & 0.019235 & RR0 & \\ 
V145 & 0.514456 & 0.514456 & 0 & 920.757864 & 0.733178 & 0.072504 & RR0 & \\ 
V146 & 0.502149 & 0.596745 & 1 & 920.939100 & 0.011152 & 0.934490 & RR0 & \\ 
V147 & 0.346495 & 0.346479 & 0 & 921.092130 & 0.000029 & 0.966839 & RR1 & \\ 
V148 & 0.467252 & 0.467290 & 0 & 920.988996 & 0.984154 & 0.016907 & RR0 & \\ 
V149 & 0.548252 & 0.548150 & 0 & 920.971600 & 0.966512 & 0.221876 & RR0 & \\ 
V150 & 0.523919 & 0.523919 & 0 & 920.811480 & 0.070358 & 0.156440 & RR0 & Bl \\ 
V151 & 0.516497 & 0.517767 & 1 & 921.141115 & 0.975617 & 0.018757 & RR0 & \\ 
V152 & 0.326130 & 0.326135 & 0 & 920.865910 & 0.017171 & 0.964094 & RR1 & \\ 
V156 & 0.532003 & 0.531987 & 0 & 921.113154 & 0.630164 & 0.956626 & RR0 & \\ 
V157 & 0.543086 & 0.542820 & 1 & 921.236948 & 0.064435 & 0.043076 & RR0 & \\ 
V159 & 0.534461 & 0.533260 & 1 & 921.009442 & 0.211989 & 0.118682 & RR0 & Bl? \\ 
V160 & 0.656928 & 0.657324 & 0 & 921.116672 & 0.022377 & 0.080551 & RR0 & \\ 
V161 & 0.526380 & 0.521640 & 1 & 921.221320 & 0.955925 & 0.073217 & RR0 & \\ 
V165 & 0.483630 & 0.483630 & 0 & 920.925300 & 0.980936 & 0.278767 & RR0 & \\ 
V168 & 0.274279 & 0.275970 & 1 & 920.865384 & \nodata & 0.849482 & RR1 & \\ 
V170 & 0.432339 & 0.432260 & 0 & 921.088722 & 0.871899 & 0.978309 & RR0 & \\ 
V171 & 0.303307 & 0.303828 & 1 & 920.786330 & 0.965052 & 0.026376 & RR1 & \\ 
V172 & 0.542261 & 0.542288 & 0 & 920.859959 & 0.997274 & \nodata & RR0 & \\ 
V173 & 0.607016 & 0.607016 & 0 & 921.186984 & 0.020296 & 0.103780 & RR0 & \\ 
V174 & 0.592278 & 0.587030 & 1 & 920.802710 & 0.829377 & 0.863071 & RR0 & Bl \\ 
V175 & 0.569703 & 0.569700 & 0 & 921.105385 & 0.958908 & \nodata & RR0 & \\ 
V176 & 0.539702 & 0.540593 & 1 & 921.105838 & 0.066155 & \nodata & RR0 & Bl? \\ 
V177 & 0.348338 & 0.348749 & 1 & 921.021524 & 0.931721 & 0.018993 & RR1 & \\ 
V178 & 0.266981 & 0.267387 & 1 & 920.748500 & 0.061053 & 0.025845 & RR1 & \\ 
V180 & 0.609131 & 0.615930 & 1 & 921.312176 & 0.010697 & \nodata & RR0 & \\ 
V181 & 0.664073 & 0.663840 & 0 & 920.813281 & 0.075114 & \nodata & RR0 & \\ 
V184 & 0.531176 & 0.532130 & 1 & 920.896900 & 0.852682 & 0.994819 & RR0 & \\ 
V187 & 0.585874 & 0.584126 & 1 & 921.311214 & 0.015553 & 0.952741 & RR0 & \\ 
V188 & 0.266510 & 0.266288 & 1 & 920.857460 & 0.983828 & 0.029642 & RR1 & \\ 
V189 & 0.612931 & 0.612940 & 0 & 920.984800 & 0.888405 & 0.869643 & RR0 & \\ 
V190 & 0.522735 & 0.522803 & 0 & 921.165465 & 0.639846 & 0.960018 & RR0 & \\ 
V191 & 0.525798 & 0.520030 & 1 & 920.770800 & 0.962122 & 0.308864 & RR0 & \\ 
V192 & 0.497111 & 0.497900 & 1 & 920.929278 & 0.017658 & 0.692467 & RR0 & \\ 
V193 & 0.747840 & 0.747840 & 0 & 920.970360 & 0.059104 & 0.059959 & RR0 & Bl? \\ 
V194 & 0.489200 & 0.489200 & 0 & 920.889000 & 0.848937 & 0.229967 & RR0 & \\ 
V195 & 0.643940 & 0.643940 & 0 & 920.898380 & 0.133118 & \nodata & RR0 & Bl? \\ 
V197 & 0.499904 & 0.499904 & 0 & 920.764676 & 0.016379 & 0.016003 & RR0 & \\ 
V200 & 0.358013 & 0.359820 & 1 & 921.002184 & 0.970811 & \nodata & RR01 & \\ 
V201 & 0.540571 & 0.541410 & 1 & 920.749258 & 0.917894 & 0.842581 & RR0 & Bl? \\ 
V202 & 0.773562 & 0.773562 & 0 & 921.023538 & 0.999889 & 0.074996 & RR0 & \\ 
V207 & 0.345329 & 0.344580 & 1 & 920.861604 & 0.042568 & 0.977804 & RR1 & \\ 
V208 & 0.338234 & 0.338020 & 0 & 920.898494 & 0.995867 & 0.961275 & RR1 & \\ 
V212 & 0.542136 & 0.542196 & 0 & 920.804712 & 0.977128 & \nodata & RR0 & Bl? \\ 
V213 & 0.299944 & 0.299706 & 1 & 920.865036 & 0.002547 & 0.904996 & RR1 & \\ 
V214 & 0.539302 & 0.540450 & 1 & 921.266446 & 0.990758 & 0.232990 & RR0 & \\ 
V214 & 0.539489 & 0.540450 & 1 & 921.227229 & \nodata & \nodata & RR0 & \\ 
V215 & 0.528949 & 0.527100 & 1 & 921.184451 & 0.835527 & 0.892058 & RR0 & \\ 
V216 & 0.346484 & 0.346484 & 0 & 920.992228 & 0.956627 & 0.030443 & RR1 & \\ 
V218 & 0.544870 & 0.544870 & 0 & 920.770010 & 0.959623 & 0.864848 & RR0 & Bl \\ 
V220 & 0.600110 & 0.600110 & 0 & 920.975090 & 0.073636 & 0.096449 & RR0 & \\ 
V221 & 0.378752 & 0.378752 & 0 & 920.843884 & 0.118019 & 0.087656 & RR1 & \\ 
V223 & 0.329210 & 0.329586 & 1 & 920.855640 & 0.894201 & \nodata & RR1 & \\ 
V229 & 0.503635 & 0.687700 & 1 & 920.978435 & \nodata & 0.974515 & RR0 & \\ 
V234 & 0.507925 & 0.549500 & 1 & 920.785750 & 0.922823 & 0.900182 & RR0 & \\ 
V235 & 0.759901 & 0.765190 & 1 & 920.924695 & 0.978616 & 0.039742 & RR0 & Bl? \\ 
V237 & 0.041984 & \nodata & 1 & 920.693148 & 0.209223 & 0.111185 & SXP & mp \\ 
V239 & 0.497584 & 0.669490 & 1 & 921.034288 & 0.080003 & \nodata & RR0 & \\ 
V240 & 0.276010 & 0.276010 & 0 & 921.004000 & 0.071012 & 0.135394 & RR1 & \\ 
V241 & 0.596172 & 0.594080 & 1 & 920.756216 & 0.766980 & 0.946418 & RR0 & \\ 
V243 & 0.634288 & 0.629910 & 1 & 921.108612 & 0.925611 & 0.947210 & RR0 & Bl? \\ 
V245 & 0.284037 & 0.284030 & 0 & 920.872941 & 0.858371 & \nodata & RR1 & \\ 
V246 & 0.339171 & 0.339140 & 0 & 920.993319 & 0.032376 & \nodata & RR1 & \\ 
V247 & 0.605350 & 0.605350 & 0 & 921.152200 & 0.010209 & 0.031635 & RR0 & \\ 
V249 & 0.533010 & 0.533010 & 0 & 921.201700 & 0.985666 & \nodata & RR0 & Bl? \\ 
V250 & 0.590310 & 0.590310 & 0 & 920.847190 & 0.823415 & \nodata & RR0 & Bl \\ 
V252 & 0.501550 & 0.501550 & 0 & 920.950300 & 0.855647 & 0.832519 & RR01 & \\ 
V253 & 0.332661 & 0.331610 & 1 & 920.850651 & 0.965971 & 0.931402 & RR01 & \\ 
V254 & 0.605519 & 0.604550 & 0 & 921.140032 & 0.873781 & \nodata & RR0 & \\ 
V255 & 0.572640 & 0.572640 & 0 & 921.203710 & 0.651841 & \nodata & RR0 & \\ 
V256 & 0.320459 & 0.318360 & 1 & 920.887774 & 0.945157 & 0.909177 & RR1 & \\ 
V258 & 0.713374 & 0.713409 & 0 & 921.050404 & 0.897383 & 0.981948 & RR0 & \\ 
V259 & 0.333463 & 0.332870 & 1 & 920.823344 & 0.037054 & 0.092532 & RR1 & \\ 
V261 & 0.444942 & 0.444140 & 1 & 920.945216 & 0.988843 & 0.972994 & RR1 & \\ 
V264 & 0.356490 & 0.356490 & 0 & 920.978930 & 0.903980 & \nodata & RR1 & \\ 
V269 & 0.355657 & 0.290000 & 1 & 921.053598 & 0.902282 & 0.952330 & RR1 & \\ 
V270 & 0.690170 & 0.690170 & 0 & 921.052730 & 0.249736 & 0.178637 & RR01 & \\ 
V271 & 0.632472 & 0.630430 & 1 & 921.344022 & 0.004165 & \nodata & RR0 & \\ 
NV286 & \nodata & \nodata & 0 & \nodata & \nodata & \nodata & N/A & \\ 
NV287 & \nodata & \nodata & 0 & \nodata & \nodata & \nodata & N/A & \\ 
NV288 & 0.036600 & \nodata & 1 & 920.682300 & 0.147541 & 0.874317 & SXP & mp \\ 
NV289 & 0.035789 & \nodata & 1 & 920.703144 & 0.045321 & 0.296544 & SXP & mp \\ 
NV290 & 0.240425 & \nodata & 1 & 920.904200 & \nodata & \nodata & RR01 & \\ 
NV291 & 0.071629 & \nodata & 1 & 920.770188 & 0.924779 & 0.940555 & SXP & \\ 
NV292 & 0.296579 & \nodata & 1 & 920.798310 & 0.887012 & \nodata & RR1 & \\ 
NV293 & 0.029462 & \nodata & 1 & 920.771768 & \nodata & \nodata & SXP & \\ 
NV294 & 0.038827 & \nodata & 1 & 920.781886 & \nodata & \nodata & SXP & \\ 
NV295 & 0.036081 & \nodata & 1 & 920.763793 & 0.842549 & 0.853940 & SXP & \\ 
NV296 & 0.445955 & \nodata & 1 & 921.127710 & 0.006637 & 0.950825 & EB & \\ 
NV297 & 0.402037 & \nodata & 1 & 920.948141 & 0.146715 & 0.748595 & RR1 & \\ 
\enddata
\tablecomments{ Full discussion of the column headings is presenting in \S3 of the text.}
\label{tab:tab3}
\end{deluxetable}

\end{document}